\documentclass[a4paper,11pt]{article}
\pdfoutput=1 

\usepackage{jheppub} %
\usepackage[T1]{fontenc}
\usepackage[utf8]{inputenc}
\usepackage{lmodern}
\usepackage{amsmath}
\usepackage{cancel}
\usepackage{bm}
\usepackage{xcolor}
\usepackage{graphicx}
\usepackage{subcaption}
\usepackage{slashed}
\usepackage{amssymb}

\def\ca{C_{\rm A}}
\def\cf{C_{\rm F}}
\def\da{d_{\rm A}}
\def\df{d_{\rm F}}

\allowdisplaybreaks
\newcommand{\co}{{\cos\theta}}
\newcommand{\p}{{\bm{p}}}

\newcommand{\q}{{\bm{q}}}

\newcommand{\norm}[1]{\frac{d^3 #1}{(2\pi)^3}}

\newcommand{\eqn}[1]{Eq.~\eqref{#1}}

\newcommand{\nn}{\nonumber\\ }
\def\be{\begin{eqnarray*}}
\def\ee{\end{eqnarray*}}
\def\beq{\begin{eqnarray}}
\def\eeq{\end{eqnarray}}
\def\rmd{{\rm d}}

\title{Jet thermalization in QCD kinetic theory}
\abstract{
We perform numerical studies in the framework of QCD kinetic theory to investigate the energy and angular profiles of a high energy parton - as a proxy for a jet produced in heavy ion collisions -  passing through a Quark-Gluon Plasma (QGP).
We find that the fast parton loses energy to the plasma mainly via a radiative turbulent quark and gluon cascade that transports energy locally from the jet down to the temperature scale where dissipation takes place.
In this first stage of the system time evolution, the angular structure of the turbulent cascade is found to be relatively collimated.
However, when the lost energy reaches the plasma temperature it is rapidly transported to large angles w.r.t. the jet axis and thermalizes.
We investigate the contribution of the soft jet constituents to the total jet energy.
We show that for jet opening angles of about 0.3 rad or smaller, the effect is negligible.
Conversely, larger opening angles become more and more sensitive to the thermal component of the jet and thus to medium response.
Our result showcases the importance of the jet cone size in mitigating or enhancing the details of dissipation in jet quenching observables. 
    }

\author[a,b]{Yacine Mehtar-Tani}
\emailAdd{mehtartani@bnl.gov}
\affiliation[a]{Physics Department, Brookhaven National Laboratory, Upton, NY 11973, USA}
\affiliation[b]{RIKEN BNL Research Center, Brookhaven National Laboratory, Upton, NY 11973, USA}

\author[c]{Soeren Schlichting}
\emailAdd{sschlichting@physik.uni-bielefeld.de}
\author[c,d]{Ismail Soudi}
\emailAdd{ismail.soudi@wayne.edu}

\affiliation[c]{Fakult\"at f\"ur Physik, Universit\"at Bielefeld, D-33615 Bielefeld, Germany}
\affiliation[d]{Department of Physics and Astronomy, Wayne State University, Detroit, MI 48201.}

\keywords{Perturbative QCD, Jet quenching, Kinetic theory}

\date{\today}

\begin{document}

\maketitle

\section{Introduction}


Understanding how QCD jets produced in heavy ion collisions degrade their energy in the presence of the Quark-Gluon Plasma (QGP) due to strong final state interactions is one of the key theoretical and experimental challenges of the heavy ion programs at RHIC and LHC. Not only do they constitute unique probes of the transport properties of the quark gluon plasma, but they also provide valuable information on non-equilibrium dynamics and the approach to thermal equilibrium of non-Abelian plasmas.  

Jet evolution in a QCD medium has been extensively studied in the past decade to extend the picture of leading radiative parton energy loss to jets as multi-partonic systems. The theoretical foundations of the former were laid in the late 1990's, by Zakharov, Baier, Dokshitzer, Mueller, Peigné, Schiff, \cite{Baier:1996kr,Zakharov:1996fv,Baier:1996sk,Zakharov:1997uu,Zakharov:1998sv,Baier:2000mf,Baier:2001yt} who showed that not only can an energetic parton passing through dense QCD matter lose energy elastically but also by medium-induced gluon radiation triggered coherently by multiple scatterings during the quantum mechanical formation time of the fluctuation. The resulting radiative spectrum is suppressed at high gluon frequency due to the so-called Landau-Pomeranchuk-Migdal effect \cite{Migdal:1956tc} as compared to incoherent Bethe-Heitler radiation \cite{Bethe:1934za}.  As a result, the average energy loss was shown to scale quadratically with medium size in contrast with the linear growth of elastic energy loss and therefore becomes the dominant energy loss mechanisms for large QCD media. Because the jet spectra are steeply falling as a function of the transverse momentum $p_T$ of the jet, jet suppression as measured by the nuclear modification factor $R_{AA}$ is sensitive to multiple soft gluon radiation that are produced copiously over the extent of the medium length. Hence, multiple radiation can be assumed to be independent in a first approximation, allowing for a Markovian description of the primary gluon radiation \cite{Baier:1996kr,Blaizot:2013vha}. More generally, the full medium-induced parton cascade can be studied in the framework of kinetic theory, which enables an end-to-end description of jet thermalization in the QGP \cite{Schenke:2009gb,Kutak:2018dim,Mehtar-Tani:2018zba,Adhya:2019qse,Blanco:2020uzy,Schlichting:2020lef,Blanco:2021usa}.

Clearly, the main limitation of the kinetic description for the medium-induced parton cascade is the fact that it does not account for the collinear QCD shower that forms as a result of the high virtuality of the initial parton associated with the hard scattering process at the origin of the jet production. The probability for such a branching process is enhanced logarithmically near the hard vertex and thus, most of the perturbative part of the collinear QCD cascade occurs early before final state interactions take place. The junction between these parton cascades of two kinds was poorly understood up until recently. Many models and Monte Carlo event generators available today use {\it ad hoc} prescriptions in order to combine the collinear virtual cascade with the medium induced one, introducing substantial theory uncertainties. The difficulty lies in the fact that the branching process is of quantum nature, and therefore one may reasonably expect quantum interference to play a crucial role in the build up of the parton shower. Some progress has been made in this direction recently in a series of works that investigated the radiative pattern of a system of two color correlated partons, a.k.a., a parton antenna \cite{Mehtar-Tani:2011vlz,Mehtar-Tani:2011lic,Armesto:2011kh,Mehtar-Tani:2012mfa,Calvo:2014cba,Moldes:2015zya,Mehtar-Tani:2021fud}. Instead of propagating independently in the plasma, if the two final state parton system is collimated enough such that its typical transverse size inside the plasma is smaller than the medium resolution scale the di-parton system remains in a color coherent state and thus behaves effectively as a single color charge. As a result, the two-parton antenna loses energy as a single parton that carries the total charge of the system. The generalization  to an arbitrary number of partons in the collinear shower was worked out in the so-called leading logarithmic approximation, where color coherence constrains the phase space for independent partons in the jet \cite{Mehtar-Tani:2017web,Mehtar-Tani:2017ypq,Caucal:2018dla}. 

Given these complications, the goal of this work is not to provide a comprehensive study of all aspects of jet evolution in the plasma in view of data comparison. Therefore, we won't touch on the initial collinear cascade and instead focus on dissecting the medium induced cascade initiated by a single parton to gain more insight onto this phase of jet evolution and provide guidance for future phenomenological studies. 

Our motivation is twofold. In addition to studying in detail the 3D structure of the medium-induced cascade, which includes energy and angular dependence of the jet constituents, we aim at gauging the sensitivity of jet evolution in the plasma to the soft medium scale in particular to the underlying dissipative processes, by dissipation we mean the irreversible process of depositing energy and momentum to the soft thermal bath.  Indeed, jet quenching observables are typically computed assuming that the medium induced cascade is perturbative and that medium properties are fully encoded in transport coefficients such as $\hat q $, which are then fitted to the data. However, it is not a given that this is a pertinent approximation at RHIC and LHC energies, where the details of energy dissipation may have a quantitative impact on seemingly perturatively computable observables. This question is reflected to a large extent by the ongoing discussion on the impact of medium response or back reaction on jet quenching observables \cite{Casalderrey-Solana:2004fdk,Renk:2006mv,Qin:2009uh,Li:2010ts,Tachibana:2014lja,Tachibana:2017syd,Casalderrey-Solana:2020rsj,Schlichting:2020lef,Yang:2022nei,Pablos:2022piv}. It is therefore necessary to identify the domain of validity of systematic perturbative computations for jet observables for precision phenomenology in the future. 

The in-medium parton cascade that resums to all orders medium induced branchings was analyzed analytically under certain simplifying assumptions and approximations \cite{Blaizot:2012fh,Blaizot:2013hx,Blaizot:2015jea}. A remarkable feature of this cascade is the emergent turbulent nature of energy transport from the leading parton energy to the infrared. In a wide range of scales connecting the jet energy typically of the order of 100-500 GeV at LHC and the plasma temperature $T\simeq 0.5$ GeV, known as the inertial window, energy is transported with constant flux without any accumulation. This feature is the hallmark of turbulence, whether it is due to quantum mechanical splittings or related to classical nonlinear wave systems. In the inertial range of frequencies, this process is characterized by self-similar dynamics encoded in universal exponents; it is local in frequency space, that is, it does not depend on the injection scale $p_T$ nor the dissipation scale $T$.  This represents an efficient mechanism for energy transport from large to low frequencies. 

In this work we shall go beyond previous works by investigating the fate of the jet four-momentum across the turbulent phase as well as the thermalization processes in the framework of kinetic theory, which allows us to address simultaneously the dynamics of parton cascade as well as the induced medium response. By including the angular dimension of the turbulent cascade into the effective kinetic description, we will in particular address the question, how energy is efficiently transferred from small to large angles w.r.t. to the jet axis.

The paper is organized as follows. In Sec. \ref{sec:Setup}, we present the linearized kinetic description and define the evolution equations which will be used to study energy loss and thermalization of high-energy partons. Sec. \ref{sec:EnergyLoss} provides an overview of our results for the energy loss and equilibration of a high energy parton in a thermal medium. Subsequently, in Sec.~\ref{sec:Soft} and \ref{sec:Stopping}, we investigate the sensitivity of out-of-cone energy loss to the equilibration of soft fragments, and the dependence on the energy scale of the hard parton. We illustrate in Sec.~\ref{sec:Quench}, how our results in QCD kinetic theory can be employed in phenomenological studies of parton and jet quenching. We summarize our most important findings and conclude in Sec.~\ref{sec:Conclusion}. 
\section{Setup -- Linearized kinetic equations}\label{sec:Setup}
\begin{figure}
    \centering
    \includegraphics[width=\textwidth]{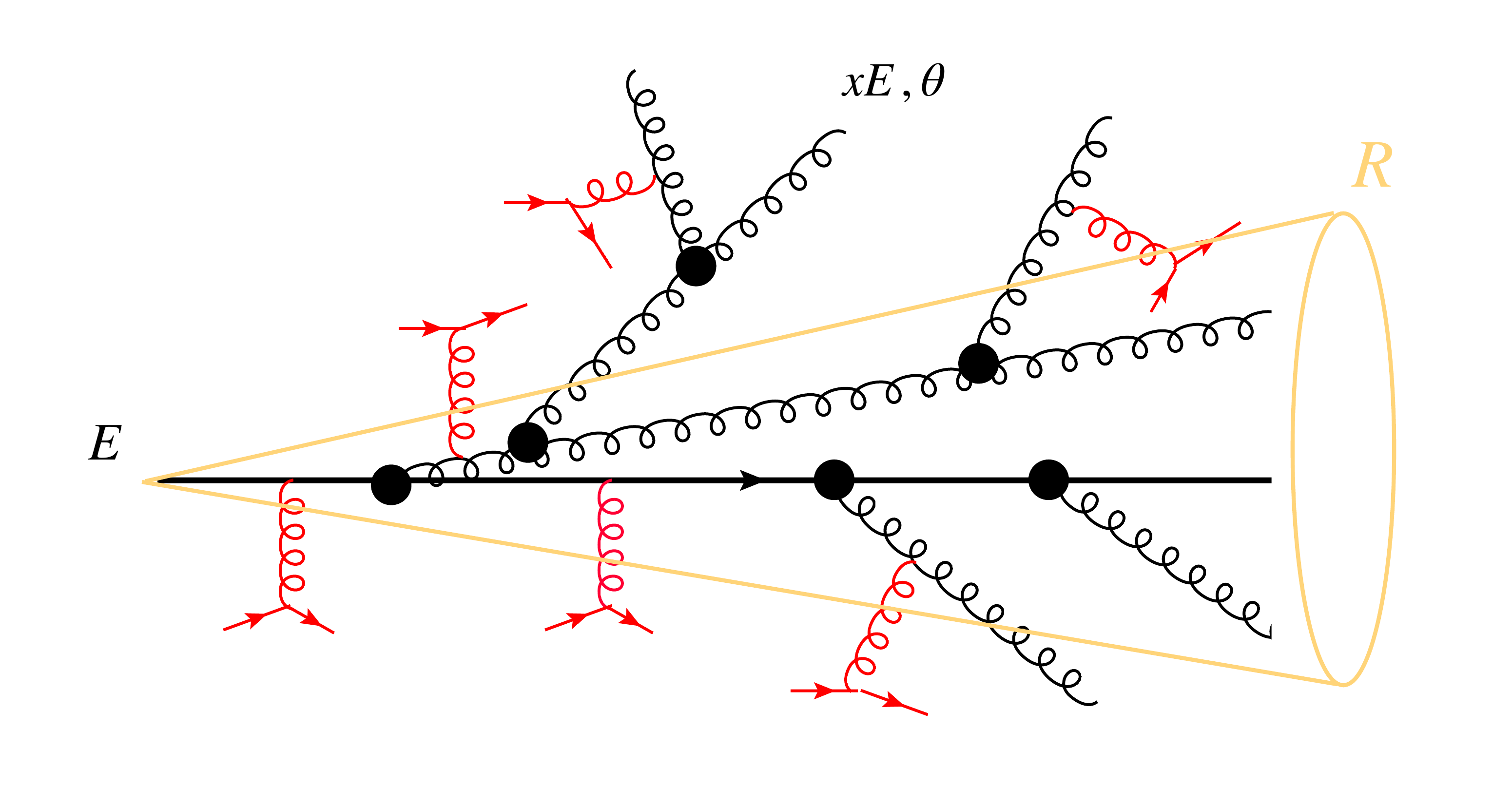}
    \caption{Schematic representation of the in-medium cascade of a hard parton produced by a hard scattering, which subsequently interacts with the QGP. We consider an initial hard parton with momentum along the $z$ direction which loses its energy due to elastic scatterings with the medium and collinear radiation. The energy distribution $D(x,\theta)$ represents the average energy and momentum distributions of the fragments originating from the initial hard parton. In this work, we investigate the average energy lost out of a jet cone of size $R$ as a result of the cascading process induced by interactions with the plasma constituents, depicted in red. Due to energy-momentum conservation, recoil effects may also contribute to the in-cone energy of the jet. }
    \label{fig:Jet-EKT}
\end{figure}

We are interested in the problem of energy deposition of high energy partons in a thermal QCD medium. We will neglect collinear branchings that may be triggered by a nascent parton with large virtuality, hence, we only consider initially on-shell quarks or gluons. Of course, a complete treatment of jet evolution in heavy ion collisions must include the latter parton cascade that forms in vacuum as well. We leave this formidable problem for a future work and focus here on the problem of energy loss and medium-induced fragmentation of a single high energy parton passing through the Quark-Gluon plasma (QGP), as illustrated schematically in Fig.\ref{fig:Jet-EKT}. 

Owing to the fact that in a typical event the elementary elastic and inelastic processes are local, i.e., their characteristic time scales are much shorter than the system size, we may use kinetic theory to describe the evolution of highly energetic partons traversing the QGP. Within this study, we will describe the QGP as a weakly coupled system in thermal equilibrium at a constant temperature $T$ (independent of spacial coordinates or time). Both the thermal QGP as well as the hard partons are then described in terms of on-shell particles by the phase-space distribution $f_a(\p,\mathbf{x},t)$ where $a=g,q_f,\bar{q}_f$ labels gluons ($g$) and quarks/anti-quarks ($q_{f}/\bar{q}_{f}$ of a given flavor $f$)\footnote{We consider $N_f=3$ degenerate flavors of massless quarks.}. In this framework, the evolution of the systems is described by the leading order QCD Boltzmann equation for the phase-space distributions $f_{a}$, which includes both the number conserving $2\leftrightarrow2$ processes as well as effective collinear $1\leftrightarrow2$ processes, and reads \cite{Arnold:2002zm,Arnold:2002ja,Ghiglieri:2015ala,Du:2020dvp}
\begin{equation} \label{eq:NL-kinetic}
    \left( \partial_t + \frac{\p}{|\p|} \cdot \mathbf{\nabla}_x \right) f_a(\p,\mathbf{x},t) = -C_a^{2\leftrightarrow2}[\{f_i\}]-C_a^{1\leftrightarrow2}[\{f_i\}]\;.
\end{equation}
In Eq.~(\ref{eq:NL-kinetic}), the elastic collision rates at leading order in the QCD coupling constant are given by \cite{Arnold:2002zm,Arnold:2002ja,Ghiglieri:2015ala,Du:2020dvp} 
\begin{align}
    &C_a^{2\leftrightarrow 2}[\{f_i\}] =  \frac{1}{4|\p_1|\nu_a} \sum_{bcd} \int_{\p_2\p_3\p_4}
        \left|{\cal M}^{ab}_{cd}(\p,\p_2;\p_3,\p_4)\strut\right|^2 (2\pi)^4 \delta^{(4)}(P + P_2 -P_3-P_4) \nonumber\\
        & \times \big\{ 
        f_a(\p)f_b(\p_2) (1\pm f_c(\p_3)(1\pm f_d(\p_4))
        - f_c(\p_3)f_d(\p_4)(1\pm f_a(\p) ) (1\pm f_b(\p_2))
        \big\}\;,
\end{align}
where ${\cal M}^{ab}_{cd}(\p_1,\p_2;\p_3,\p_4)$ stand for the relevant spin and color averaged two-to-two tree-level matrix elements  (see. e.g. \cite{Arnold:2002zm}), while $\nu_{g}=2d_A=2(N_c^2-1)$ and $\nu_{q_f}=\nu_{\bar{q}_f}=2d_{F}=2N_c$ denote the color and spin degeneracy factors of gluons and quarks. We note that in a thermal QGP the infrared divergencies of the matrix elements  ${\cal M}^{ab}_{cd}(\p_1,\p_2;\p_3,\p_4)$ associated with small momentum exchange in the $t$ and $u$ channels at tree-level are regularized by screening effects that can be accounted for by using HTL propagators as discussed in Refs.~ \cite{Arnold:2002zm,Arnold:2002ja,Arnold:2003zc,Ghiglieri:2015ala}. We will follow the approach of Ref.~\cite{Arnold:2003zc} and refer to ~App.~\ref{ap:HTL} for additional details.

Since each gluon radiation comes at the expense of an additional power of the coupling constant, inelastic processes are naively suppressed in plain perturbation theory. However, this suppression is compensated by an enhancement of (nearly) collinear splittings, such that even at leading order these processes cannot be neglected. Since nearly collinear splittings can be induced by coherent multiple splittings over the course of the formation time, in fact it becomes necessary to resum their effects into an effective collinear rate $\frac{d\Gamma^a_{bc}(\p,z)}{dz}$\footnote{We follow the notation of \cite{Arnold:2008iy}, where the rate $\frac{d\Gamma^a_{bc}(\p,z)}{dz}$ includes a factor $\frac{1}{\nu_a}$ to account for the spin and color degeneracy factors $\nu_a$ of the parent parton $a$.}, which is integrated over the typically small transverse momentum transfer generated during the splitting, and most importantly accounts for the Landau, Pomeranchuk and Migdal (LPM) effect~\cite{Landau:1953um,Migdal:1956tc,Baier:1996kr,Baier:1996sk,Zakharov:1996fv,Zakharov:1997uu,Zakharov:1998sv}.  By following the approach by Arnold, Moore and Yaffe (AMY) \cite{Arnold:2002zm,Arnold:2002ja}, to perform the re-summation of said multiple scatterings, one obtains the collision integral for  such effective $1 \leftrightarrow 2$ collinear processes which reads  \cite{Arnold:2002zm,Arnold:2002ja,Du:2020dvp}
\begin{align}
    &C_a^{1\leftrightarrow 2}[\{f_i\}]=\nonumber\\
    &\sum_{bc}\int_0^1 dz\Bigg\{\frac{1}{2} \frac{d\Gamma^a_{bc}(\p,z)}{dz}\Big[ f_a(\p)(1\pm f_b(z\p))(1\pm f_c(\bar z \p)) - f_b(z\p)f_c(\bar z \p)(1\pm f_a(\p)) \Big] - \frac{\nu_b}{\nu_a} \nonumber \\
    & \times\frac{1}{z^3}\frac{d\Gamma^b_{ac}(\frac{\p}{z}\;,z)}{dz}
    \Big[ f_b\left(\frac{\p}{z}\right)(1\pm f_a(\p))\left(1\pm f_c\left(\frac{\bar z}{z}\p\right)\right) -f_a(\p)f_c\left(\frac{\bar z}{z}\p\right)\left(1\pm f_b\left(\frac{\p}{z}\right)\right) \Big] \Bigg\}\;,
\end{align}
Here the first term in the second line describes the $1\to 2$ collinear splitting of a parent parton of momentum $p$ into two offspring of momenta $z p$ and $\bar z p = (1-z)p$, while the second term in the first line describes the reverse $2\to 1$ process. Similarly, the second term in the third line describes the merging of two partons with momenta $p$ and $\bar{z} p/z$ into a parton of momentum $p/z$, while the first term in the third line describes the reverse splitting process.

Note that, since we are interested in the evolution of the hard parton in a large medium, we shall neglect finite size effects in the calculation of the medium induced splitting rates $\frac{d\Gamma^a_{bc}(\p,z)}{dz}$.  Stated differently, we will work in the limit of a constant or saturated radiation rate $\frac{d\Gamma^a_{bc}(\p,z)}{dz}$ as computed for an infinite medium \cite{Arnold:2002zm,Arnold:2002ja,Ghiglieri:2015ala}, and we follow the procedure outlined in App.~A of \cite{Schlichting:2020lef} to calculate the associated rates. Clearly, this is a good approximation, so long as the quantum mechanical formation time $t_{f}$ of the splitting is much smaller than the medium length $L$. Specifically, for the splitting of a highly energetic parton $p \gg T$, this condition takes the form 
\begin{align}
t_{f}(\omega)\simeq \sqrt{\frac{\omega }{ \hat q }} \ll L\,.
\end{align}
where $\omega = z(1-z)p$ and $\hat q $ is the quenching parameter (see the following reviews   \cite{Mehtar-Tani:2013pia,Blaizot:2015lma} and references therein for more details). Hence, this treatment is justified whenever \begin{align}
\omega \ll  \omega_c \equiv \hat q L^2\,,
\end{align}
i.e. for very asymmetric splittings and/or for sufficiently small parent momentum $p\ll \hat{q}L^2$ in a large medium where $\hat{q}L^2 \gg T$.  What we call the quenching parameter $\hat{q}$, can be related to the mean transverse momentum squared a parton acquires due to subsequent scatterings with a medium of length $L$ as
\begin{align}\label{eq:qhat}
    \hat{q} =  \frac{\langle k_\bot^2\rangle}{L}\;.
\end{align}
Specifically, at leading order of the perturbative expansion the coefficient is given by \cite{Arnold:2002zm,Ghiglieri:2015ala}
\begin{align}
    \hat{q} = C_R\hat{\bar{q}} = \frac{g^2TC_R}{2\pi}m_D^2 \ln\frac{Q}{m_D}\;,
\end{align}
where we introduce the notation $\hat{\bar{q}}$ as the quenching parameter stripped of its color factor $C_R$ of the hard parton, while $m_D^2=g^2T^2(N_c/3 + N_f/6)$ is the leading order Debye mass and $Q$ stands for the hard scale of the process \cite{Mehtar-Tani:2019ygg, Barata:2020sav}. While for a realistic medium of length $L=5~{\rm fm}$ and a quenching factor  $\hat{q}=1~{\rm GeV}/{\rm fm}^2$, we find a typical scale $\omega \ll 25$ GeV, throughout this manuscript will extend beyond this phase-space for simplicity.

Evidently, the QCD Boltzmann equation (\ref{eq:NL-kinetic}) is a non-linear kinetic equation that, in its general form, can and has been used to study the thermalization of far-from equilibrium QCD plasmas~\cite{Kurkela:2018oqw,Du:2020dvp}. Since, we are interested in the relaxation of hard partons that are characterized by very small occupation numbers $\delta f_a(\p,\mathbf{x},t) \ll 1 $, the problem is actually simpler, as we may treat the disturbance due to the presence of the hard parton as a perturbation on top of thermal QGP. By treating the thermal QGP as static and homogeneous for simplicity, we can then decompose the phase space distribution as 
\begin{align}
f_a(\p,\mathbf{x},t) =& n_a(p) +\delta f_a(\p,\mathbf{x},t)\;,
\label{eq:Linearization}
\end{align}
where $n_a(p)$ stands for the equilibrium Bose-Einstein distribution $\left( n_g(p) = \frac{1}{e^{p/T}-1} \right)$ for gluons and respectively the Fermi-Dirac distribution $\left( n_q(p) = \frac{1}{e^{p/T}+1} \right)$ for quarks and anti-quarks. We then linearize Eq.~(\ref{eq:NL-kinetic}) w.r.t. the perturbation $\delta f_a(\p,\mathbf{x},t) \ll n_a(p)$ induced by the initial presence of a hard parton. Since the explicit form of the linearized kinetic equations turns out to be rather lengthy, we refrain from reproducing them in this section and instead refer the interested reader to App.~\ref{ap:HTL} for the implementation of the elastic collision integral, while the inelastic radiation collision integrals can readily be found in \cite{Schlichting:2020lef}. 

While previous works~\cite{Schlichting:2020lef} have employed essentially the same QCD kinetic setup\footnote{Note that in contrast to the present study, the elastic processes in ~\cite{Schlichting:2020lef} were treated in the small-angle approximation.} to study the degradation of the energy of hard partons, we will go beyond that and as illustrated in Fig.~\ref{fig:Jet-EKT} study the fully differential momentum distribution of jet fragments in energy ($xE$) and polar angle ($\theta$) w.r.t. the initial parton momentum. We emphasize, that the extension of the previous study to the angular distribution is crucial in the context of jet physics since by definition a jet is defined by a collection of particles that fall within a cone defined by an opening angle $R$. By explicitly keeping track of the inclusive momentum distribution of jet fragments, the problem of jet energy loss then reduces to investigating the mechanism by which energy carried by fragments ``leaks'' out of the jet cone as illustrated in Fig.~\ref{fig:Jet-EKT}, which is the central objective of this study.
 
Before we discuss this further, we wish to make contact with standard notations for the distribution of jet fragments in terms of the momentum fraction $z$ and the angle $\theta$ formed by the initial parton momentum $\p_{\rm in}$ and the measured final parton $\p$. By following previous works~\cite{Blaizot:2013vha,Blaizot:2014ula,Blaizot:2014rla}, the in-medium evolution of the jet shower is then conveniently studied in terms of the differential energy distribution $D_{a/jet}$ of particles $a$ inside a jet
\begin{align}
\label{eq:FragmentationFct}
D_{a/\rm jet}(x,\theta,t)\equiv& x \frac{\rmd N_a}{\rmd x \rmd\cos\theta}\nonumber\\
=&\nu_a\int d^3\mathbf{x} \int \norm{p}~\frac{|\p|}{E}~\delta\Big(\frac{|\p|}{E}-x\Big)\delta\Big(\frac{\p \cdot \hat{e}_z}{x E}-\cos\theta \Big) \delta f_a(\p,\mathbf{x},t)\;,
\end{align}
where $E$ is the total jet energy and $\hat{e}_z = \frac{\p_{\rm in}}{|\p_{\rm in}|}$ is the unit vector along the initial parton direction, and we have integrated out the azimuthal angle $\phi$ as well as the spatial coordinate $\mathbf{x}$. By integrating over polar angle $\cos(\theta)$, we then recover the energy distribution studied in earlier works \cite{Kutak:2018dim,Mehtar-Tani:2018zba,Adhya:2019qse,Blanco:2020uzy,Schlichting:2020lef,Blanco:2021usa} 
\beq 
D_{a/\rm jet}(x,t) = \int_{-1}^{1} \rmd \cos(\theta)~ D_{a/\rm jet}(x,\theta,t)\,,
\eeq
which as a result of energy conservation is normalized to unity, i.e.
\beq 
\label{eq:NormD}
\sum_{a} \int_0^{\infty} \rmd x \int_{-1}^{1} \rmd \cos\theta~ D_{a/\rm jet}(x,\theta,t) = \sum_{a} \int_0^{\infty} \rmd x ~D_{a/\rm jet}(x,t) =1\;,
\eeq
at any time.

\section{Energy loss and equilibration}\label{sec:EnergyLoss}

\begin{figure}
    \centering
    \includegraphics[width=0.5\textwidth]{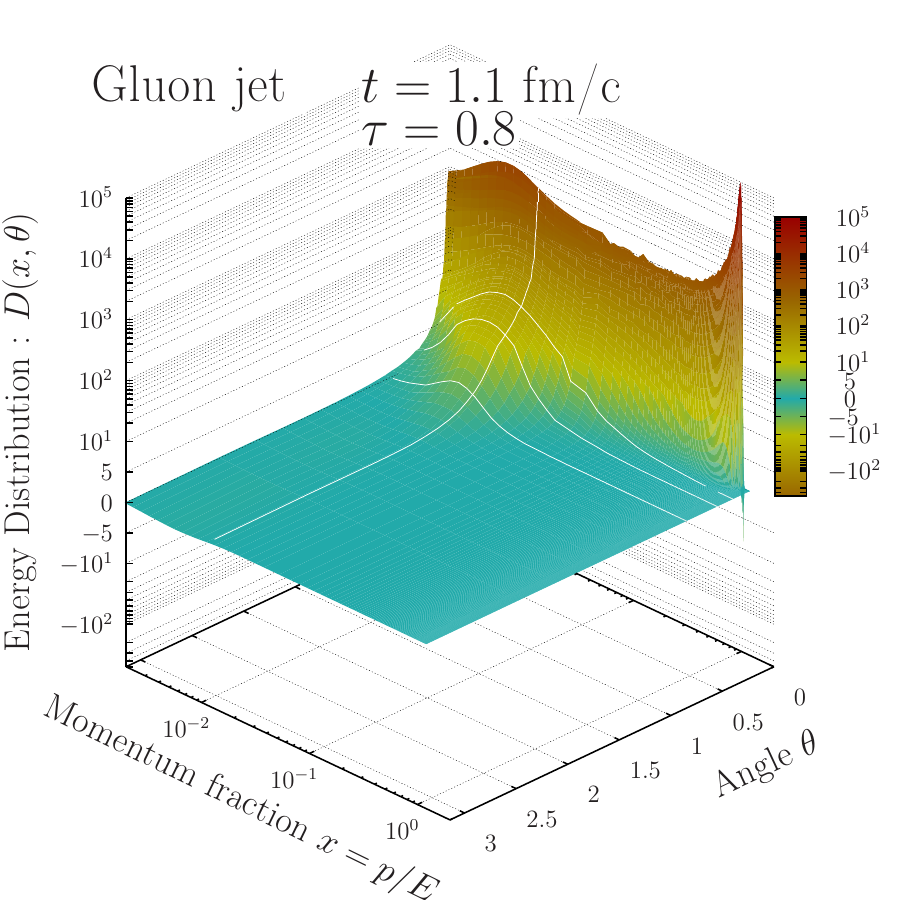}\includegraphics[width=0.5\textwidth]{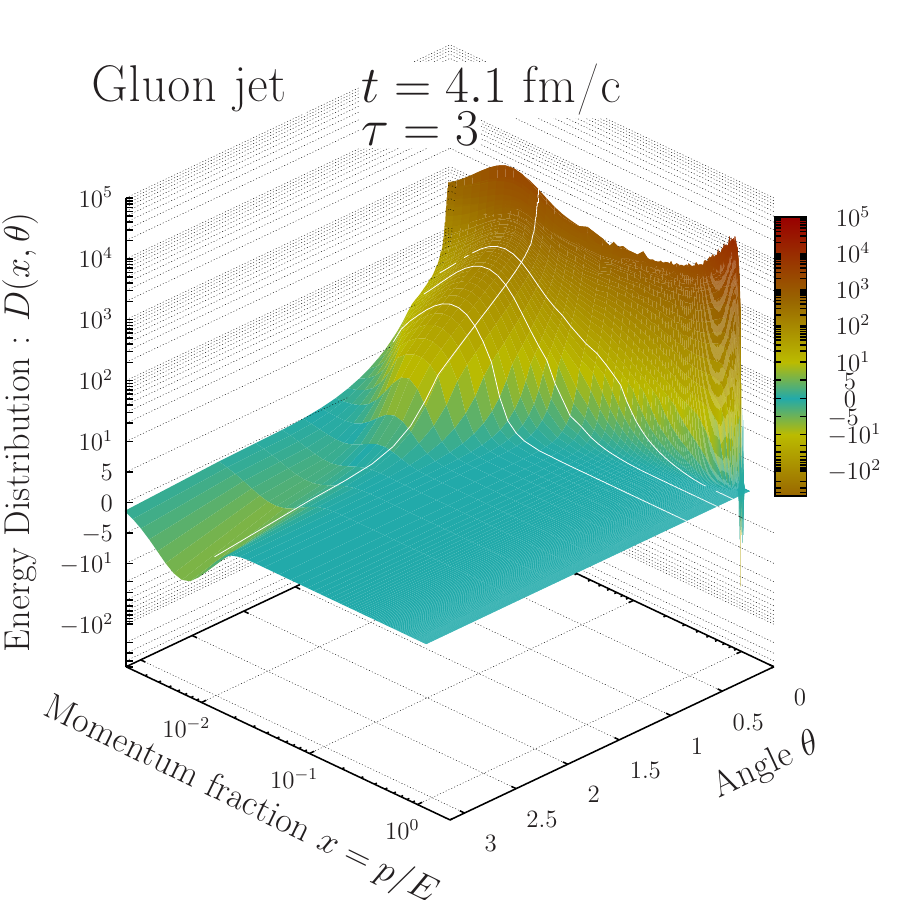}
    \includegraphics[width=0.5\textwidth]{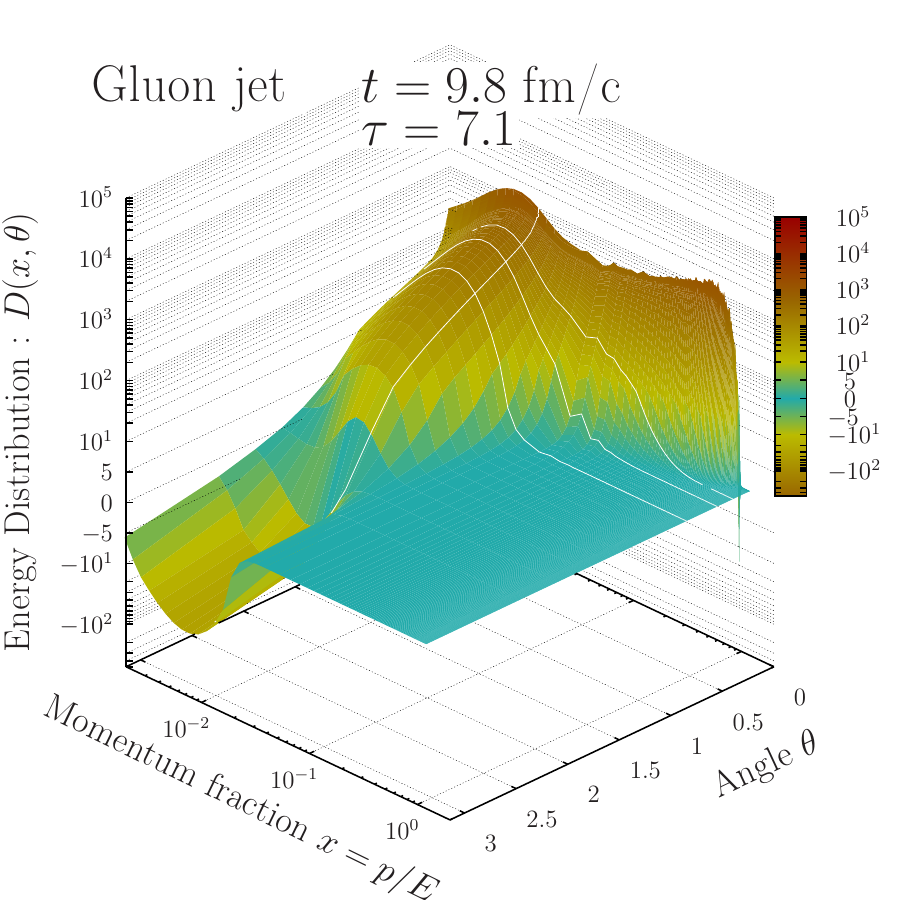}\includegraphics[width=0.5\textwidth]{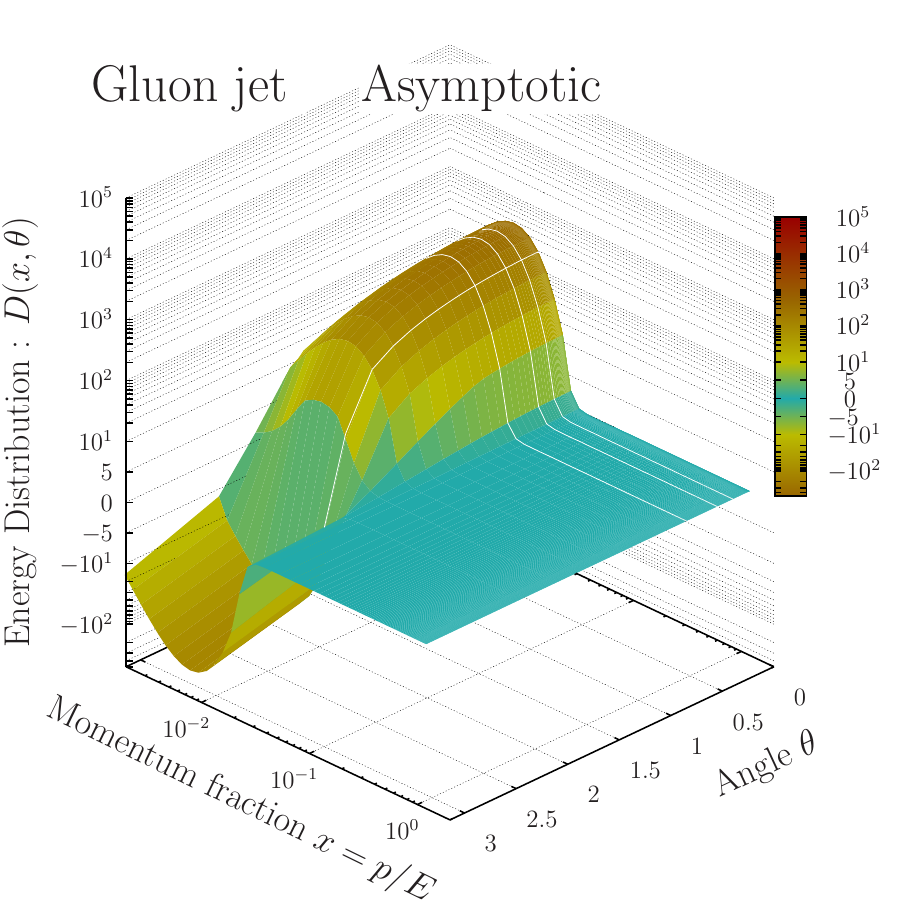}
    \caption{ Evolution of the energy distribution $D(x,\theta)=\sum_a D_{a/g-\rm jet}(x,\theta)$ of all species for a gluon jet with initial energy $E=500~T$ at different times $t=1.1,4.1,9.8$ fm/c as a function of momentum fraction $x=\frac{p}{E}$ and angle $\theta$. The bottom right panel shows the asymptotic distribution. White lines along the angle axis represent constant angles from small to large $\theta = 0.16,0.32,0.62$, while the one along the momentum fraction axis represent constant momentum fraction of $x=2\pi T/E$. }
    \label{fig:Dist3D}
\end{figure}

We are now equipped with the linearized kinetic equations that are necessary to study the problem of energy loss and equilibration of hard partons or jets inside a thermal QGP. Starting from an initial condition that represents a single hard quark or gluon moving in the $+z$ direction
\beq\label{eq:initial-cond}
D^{(0)}_{a/b-\rm jet}(x,\theta)\equiv D_{a/b-\rm jet}(x,\theta,t=0)= \delta_{ab}\delta(1-x) \, \delta(1 - \cos\theta)\,,
\eeq
we aim at a quantitative understanding of the non-equilibrium evolution of the system from the initial condition in \eqn{eq:initial-cond} all the way towards thermal equilibrium, where at asymptotically late times the only information retained from the jet is associated with the excess of the conserved energy $\delta E=E$, longitudinal momentum $\delta P_z=E$ and valence charge $\delta N_f=0$ for a gluon jet and $\delta N_f=1$ for a quark jet. 

We will consider two types of initial conditions, corresponding to a highly energetic gluon or quark, which we will refer to as $g- \rm jet$ and $q-\rm jet$ respectively. In practice, in our numerical simulations we will model the delta functions in the initial condition by a narrow Gaussian of width $\sigma =10^{-3}/\sqrt{2}$, such that e.g. for a gluon jet 
\begin{align}
    \label{eq:Angulargluon-jet1}
    &D_{g/g- \rm jet}(x,\theta)=\frac{1}{N}\exp\left\{-\frac{(1-x\cos\theta)^2 + x^2 \sin^2\theta }{2\sigma^2 } \right\}\;, \\
    &D_{q/g-\rm jet}(x,\theta)=0\;, \qquad 
    D_{\bar{q}/g-\rm jet}(x,\theta)=0\;,
    \label{eq:Angulargluon-jet2}
\end{align}
where the normalization factor is given by 
\begin{equation}
    N = \int_{-1}^{1} \rmd\co\int_{0}^{\infty} dx~ e^{-\frac{(1-x\cos\theta)^2 + x^2\sin^2\theta }{2\sigma^2 }}\;\;,
\end{equation}
such that the normalization condition in Eq.~(\ref{eq:NormD}) remains valid.
In the limit of vanishing width ($\sigma\to 0$), the parameterization of the Gaussian initial conditions is equivalent to the delta functions in Eq.~(\ref{eq:initial-cond}).
When considering quark jets, we will for simplicity assume an equal abundance of $u,d,s$ light flavor jets, such that the initial conditions for the evolution are also degenerate in flavor, as discussed in more detail in App.~\ref{ap:HTL}. If not stated otherwise, throughout this section, we will present results for the evolution of gluon jets with energy $E=500~T$ and consider $g=2$ which corresponds to a realistic coupling strength $\alpha_s=\frac{g^2}{4\pi}\simeq 0.3$ at RHIC an LHC energies. 

When studying the time evolution of the system, it is useful to recall that the medium induced cascade is characterized by a single time scale \cite{Blaizot:2012fh,Blaizot:2013hx,Blaizot:2015jea}
\begin{equation}\label{eq:SplittingTime}
t_{\rm th} \sim \frac{1}{\alpha_s } t_{f}(E) \sim \frac{1}{\alpha_s } \sqrt{\frac{E}{\hat q }}\,.
\end{equation}
Its physical meaning is that of the time it takes for a hard parton to undergo a radiative break-up and thermalize in the plasma, and it is also sometimes referred to as the stopping time. 
We may use this characteristic time scale to adopt the dimensionless time variable 
\begin{equation}\label{eq:TauScaling}
    \tau \equiv \frac{t}{t_{\rm th}}= g^4 T \sqrt{\frac{T}{E}}t\;,
\end{equation}
where we have used the fact that for a thermal plasma $\hat q \propto g^4 T^3$. Alternatively, to facilitate a connection with heavy ion phenomenology, our results will also be presented in terms of the physical time in units $\rm fm/c$ by using a thermal medium with temperature $T=200$ MeV to set the physical scale. However, we shall refrain from a detailed phenomenological study since our evolution does not include important effects such as finite size effects for the medium-induced splitting rates, the space-time evolution of the medium and the collinear vacuum cascade. 

Before delving into the detailed discussion of our numerical results, let us first highlight and summarize the generic features of the time evolution of the energy and angular distribution of the jet fragments in a 3D plot. For this purpose, the evolution of the total energy distribution of all species 
\beq
D(x,\theta,t)=\sum_a D_{a/\rm jet}(x,\theta,t)
\eeq
is presented in Fig.~\ref{fig:Dist3D} as a function of momentum fraction $x=p/E$ and polar angle $\theta$ at three different evolution times $t=1.1,4.1,9.8 {\rm fm/c}$ and compared to the asymptotic distribution in the bottom right panel of Fig.~\ref{fig:Dist3D}.\footnote{Note that we use an unconventional plotting scheme for the $z$-axis representing the distribution $D(x,\theta)$, with a linear scale from the values $D(x,\theta)=-10$ to $D(x,\theta)=10$ and a logarithmic scaling beyond for $|D(x,\theta)|>10$. The negative values of the distribution are represented as a dip in the $z$-axis below $z=0$, as opposed to conventional logarithmic scales.}

Starting at early times in the top left panel, the initial hard parton peak is clearly visible at momentum fraction $x\sim 1$ and small angles $\theta \sim 0$, however collinear radiation leads to a sizable population of the intermediate to soft sector $x < 1$ in the collinear region with no noticeable broadening to large angles.

While the hard momentum sector $x \sim 1$ continues to be depleted by collinear radiation, the soft fragments undergo a successive broadening by which the energy deposited in the soft sector $x\lesssim 2\pi T/E$ at small angles is transported to larger angles out of the jet cone $R< 1$. At the intermediate times $t=1.1, 4.1, 9.8$~fm/c the angular distribution in the soft sector thus extends all the way to large angles $\theta \sim 1$. Conversely, only a minimal broadening of the hard partons is achieved, such that only the broadening near the medium scales $x\lesssim 2\pi T/E$ contributes to the energy at large angles. 

Eventually, at late times $t=9.8$fm/c, we observe that the hard parton peak is mostly depleted and only the remnants of the energy at intermediate scales continues to equilibrate with the medium, via the same thermalization mechanism. Due to the continuous influx of energy from these semi-hard fragments, the soft sector $x \lesssim 2\pi T/E$ also remains out of equilibrium, featuring an enhancement of the energy distribution at narrow angles. Nevertheless, by this time the peak in the soft sector has broadened considerably to angles $\theta \sim 1$, as energy deposited into the soft sector at an earlier stage has had sufficient amount of time to undergo kinetic equilibration. Notably, at very large angles $\theta \gtrsim 1.3$, there is also a negative contribution to the energy distribution, which can be associated with recoiling thermal particles that due to their interaction with the nearly collinear constituents get dragged along the jet axis. While initially, this is a rather small effect, this negative contribution becomes increasingly important as more and more nearly collinear fragments $\theta \ll 1$ drag along the constituents of the thermal QGP. Stated differently, the nearly collinear cascade deposits an almost equal amount of energy $\delta E$ and longitudinal momentum $\delta P^{z}$ in the soft sector, which over the course of the thermalization of the soft sector is then redistributed among the soft fragments, leading to an increase of the number of particles moving parallel and a decrease of the number of particles moving anti-parallel to the jet axis. Such a depletion of the away side is also referred to as the diffusion wake valley and has been subject to ongoing extensive investigation using the hydrodynamic response to jet energy loss \cite{Casalderrey-Solana:2004fdk,Renk:2006mv,Qin:2009uh,Li:2010ts,Tachibana:2014lja,Tachibana:2017syd,Casalderrey-Solana:2020rsj,Yang:2022nei} (see \cite{Cao:2020wlm} for a review).

Eventually, at asymptotically late times all hard fragments have thermalized inside the medium, and the only leftover signal is due to the change of the conserved energy $\delta E$,  momentum  along the jet axis $\delta P_z$ and valence charge $\delta N_{f}$ of the jet, which lead to a change of the temperature $T$, the flow velocity $u^{z}$ and chemical potential $\mu_f$. By matching the changes in the thermodynamic variables
as detailed in App.~\ref{ap:Equilibirum},
the corresponding asymptotic energy distribution for each species can be determined as
\beq
\label{eq:LinearEquilibrium}
D_{a/jet}^{(eq)}(x,\theta)=\frac{\nu_a}{(2\pi)^2}(xE)^3\left[ \frac{1}{4} \frac{\delta E}{e(T)} T \partial_T + \frac{3}{4} \frac{\delta P^{z}}{e(T)} \partial _{u^{z}} + \frac{\delta N_f}{\chi(T)T}  \partial_{\frac{\mu_{f}}{T}} \right] \left.n_{a}\Big(p_{\mu}u^{\mu} \Big) \right|_{u^{z}=0}
\eeq
and the asymptotic energy distribution of all species is given by 
\beq
\label{eq:DeqSolMain}
D^{(eq)}_{a}(x,\theta)= \sum_{a}
\frac{\nu_a}{(2\pi)^2} \frac{(xE)^4}{4 e(T)}~\frac{E}{T}~\left[ 1+3\cos(\theta)\right]n_{a}(xE) (1\pm n_{a}(xE))
\eeq
where $\chi(T)$ and $e(T)$ are the quark flavor susceptibility and energy density as a function of temperature, and we assumed the equation of state of an ultra-relativistic ideal gas. By following the evolution up to very late times, we find that the above asymptotic distribution, which features an increase in temperature and a boost along the jet axis, is correctly recovered in our numerical simulations as can be seen in the bottom right panel of Fig.~\ref{fig:Dist3D}.  

Summarizing the results in Fig.~\ref{fig:Dist3D}, we find that the re-distribution of energy outside a jet cone is governed by a two-step scenario, where 
\begin{itemize}
    \item[1)] Hard partons with $x\gg 2\pi T/E$ remain collinear but undergo a radiative break-up due to multiple successive splittings, that deposits energy and momentum into the soft sector $x\lesssim 2\pi T/E$ via a turbulent cascade
    \item[2)] Soft fragments with $x\lesssim 2\pi T/E$ thermalize, thereby transporting the continuous influx of energy and momentum due to the collinear cascade out to large angles $\theta \sim 1$. However, due to the continuous influx of energy, the soft sector remains out of equilibrium throughout the entire evolution.
\end{itemize}
Below, we further scrutinize the evolution of the energy and angular spectra during this process and explore the consequences of this dynamics for parton and jet energy loss. 

\subsection{Momentum structure of the cascade}
\begin{figure}
    \centering
    \includegraphics[width=0.5\textwidth]{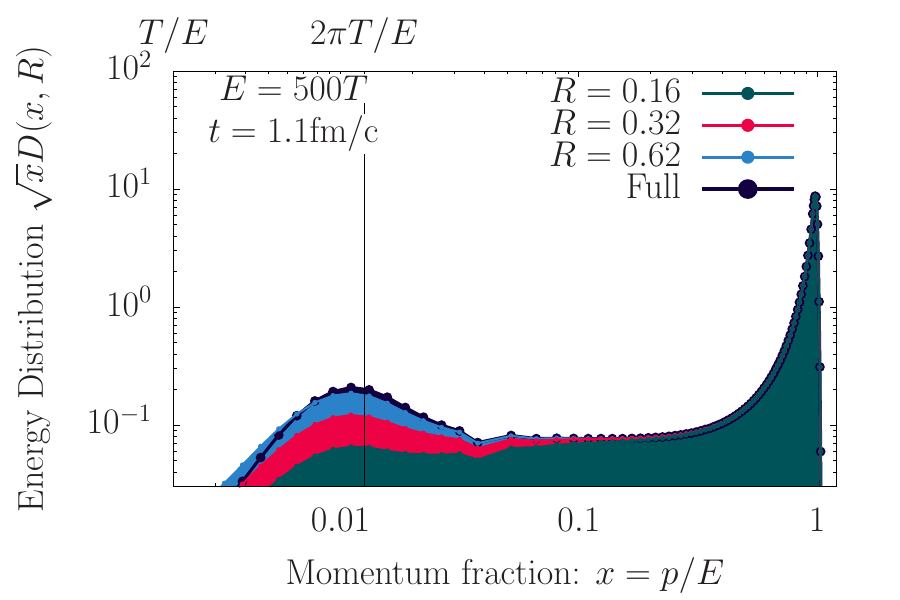}\includegraphics[width=0.5\textwidth]{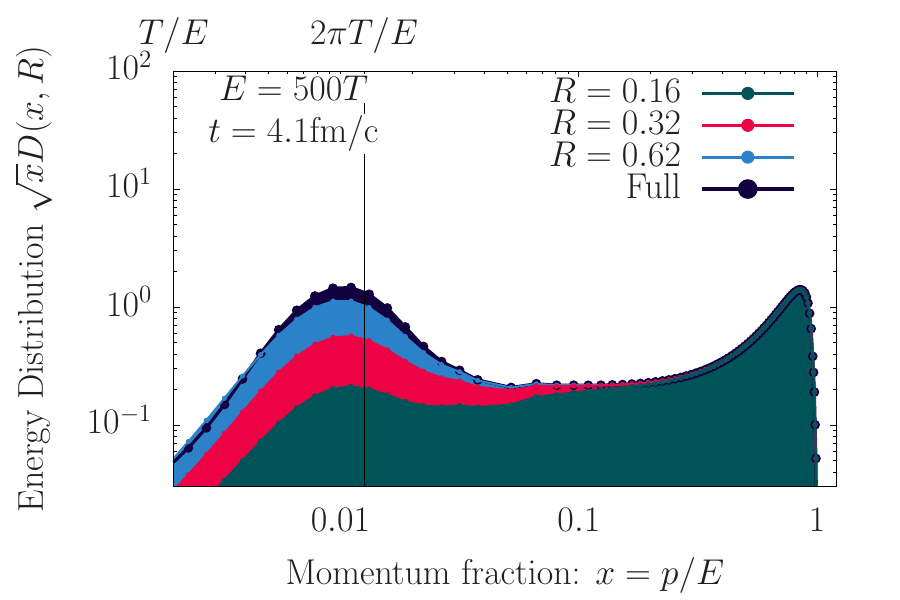}
    \includegraphics[width=0.5\textwidth]{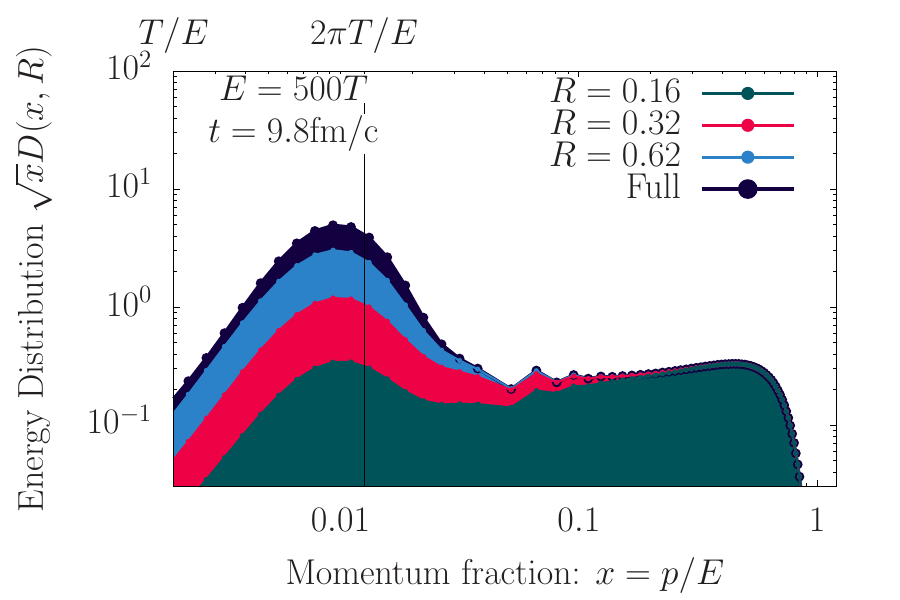}\includegraphics[width=0.5\textwidth]{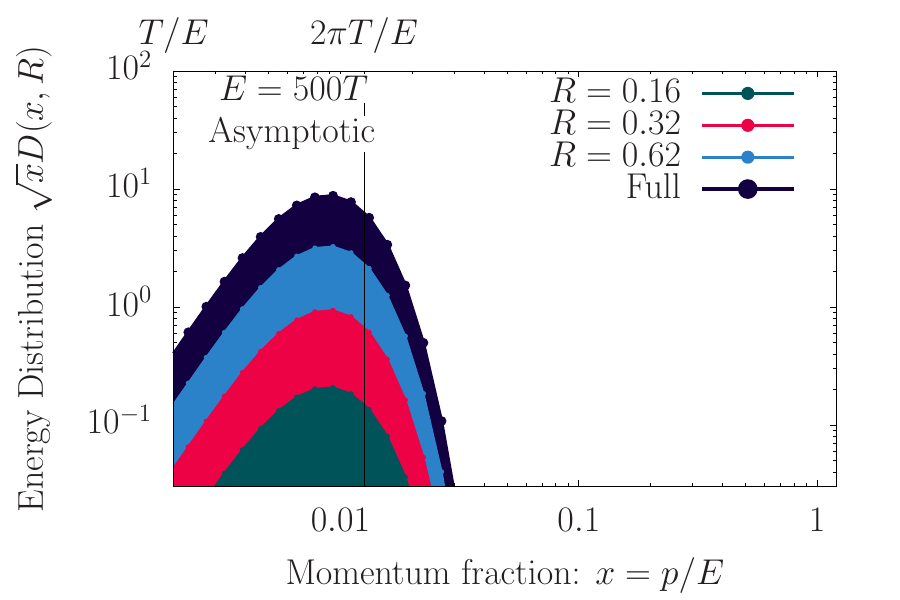}
    \caption{Evolution of the energy distribution as a function of the momentum fractions $x=\frac{p}{E}$ with a decomposition into different angular regions as described in Eq.~(\ref{eq:EnergyDist}). The shaded regions help to visualize the change in the energy distribution as function of the jet cone size $R$. }
    \label{fig:EnergyDist}
\end{figure}
Now that we have established the general picture, we will further investigate the energy spectra. In order to make explicit contact with QCD jets, we shall explore the structure of the energy distribution by considering the angular integrated energy distribution that measures the energy that remains inside a cone of opening angle $R$, 
\begin{equation}\label{eq:EnergyDist}
    D(x,R,t) = \sum_{a} \int_{\cos R}^{1} \rmd\co~ D_{a/\rm jet} (x,\theta,t)\;,
\end{equation}
where we integrate the energy distribution up to a cone of size $R$ w.r.t. the momentum of the initial hard parton.
We show in Fig.~\ref{fig:EnergyDist} the full distribution $D(x,t) \equiv D(x,R=\pi,t)$ integrated over all angles, together with the distributions $ D(x,R,t)$ for different angular cones of size $R = 0.16,0.32,0.62$ at three different times $t=1.1,4.1,9.8$~fm/c of the evolution and at asymptotically late times.

Before we turn to the discussion of our numerical results, we note that the fully integrated distribution $D(x,t)$, has been investigated in earlier kinetic evolution studies by including in-medium collinear radiation only \cite{Jeon:2003gi,Blaizot:2013hx,Blaizot:2015jea,Mehtar-Tani:2018zba} or by also including elastic scattering with the medium \cite{Iancu:2015uja,Schlichting:2020lef}.
One of the most important conclusions of these studies is  that the energy loss is driven by the radiative break-up of hard fragments due to successive in-medium splittings. Since the radiative break-up due to multiple (quasi-democratic) splittings is a self-similar process, it leads to a so-called \emph{inverse energy cascade}, i.e. a turbulent cascade that transports energy all the way from very high $(\sim E)$ to very low $(\sim T)$ momentum scales, which can be studied using the theory of weak wave turbulence \cite{Blaizot:2013hx,Blaizot:2015jea,Mehtar-Tani:2018zba,Schlichting:2020lef}. By focusing on the inertial range of intermediate momentum fractions $T/E \ll x \ll 1$ where radiative processes dominate, the evolution equations can be approximated as~\cite{Mehtar-Tani:2018zba,Schlichting:2020lef}
\begin{align}
\label{eq:simplebranching}
&\partial_{t}D_{a/\rm jet}(x,\theta,t) \simeq \nn
&\sum_{bc} \int_{0}^{1}dz~\left[ \frac{d\Gamma^{b}_{ac}}{dz}\Big(\frac{xE}{z},z\Big) D_{b/\rm jet}\Big(\frac{x}{z},\theta,t\Big)- \frac{1}{2} \frac{d\Gamma^{a}_{bc}}{dz}\Big(xE,z\Big) D_{a/\rm jet}\Big(x,\theta,t\Big) \right]\;, 
\end{align}
where we dropped all sub-leading elastic and inelastic processes, as well as all Bose enhancement and Fermi suppression factors. By employing the so called harmonic-oscillator approximation for the high-energy rates~\cite{Arnold:2008zu,Arnold:2008iy,Schlichting:2021idr}
\beq
\frac{d\Gamma^{a}_{bc}}{dz}\Big(xE,z\Big) \simeq \frac{\alpha_s}{2\pi} P_{ba}(z)~\sqrt{\frac{\hat{\bar{q}}}{z(1-z)xE}}~\sqrt{z C_{c}^{R} +(1-z)C_{b}^{R}-z(1-z)C_{a}^{R}}
\eeq
where $P_{ba}(z)$ denote the leading order QCD splitting functions and $C_{a/b/c}^{R}$ denote the Casimir of the representation of particles $a,b,c$, i.e. $C_{a}^{R}=C_{F}=\frac{N_c^2-1}{2Nc}$ for quarks $a=q_{f},\bar{q}_{f}$ and  $C_{a}^{R}=C_{A}=N_c$ for gluons $a=g$, the evolution equations reduce to the ones in \cite{Mehtar-Tani:2018zba} and the turbulent power spectrum can be determined as $D_{a/jet}(x) \propto 1/\sqrt{x}$ within the inertial range of momentum fractions $T/E \ll x \ll 1$ \cite{Blaizot:2013hx,Blaizot:2015jea,Mehtar-Tani:2018zba}. Since the in-medium splittings are concentrated in the collinear region, we can expect to recover this turbulent spectrum at intermediate momentum fractions $T/E \ll x \ll 1$ in the small angle region $\theta \ll 1$. However, it is also clear that for smaller momentum fractions $x\lesssim T/E$ the effective description in Eq.~(\ref{eq:simplebranching}) breaks down, as all kinds of elastic and inelastic processes become equally relevant for the kinetic evolution and ultimately lead to the thermalization of the soft sector.

Starting at early times ($t=1.1 {\rm fm}/c$) shown in the top left panel of Fig.~\ref{fig:EnergyDist}, the energy distribution is strongly peaked around the energy of the original hard parton ($x\sim 1$), and at intermediate momentum fractions $T/E \ll x \ll 1$ shows the expected turbulent $\sim 1/\sqrt{x}$ spectrum due to medium-induced radiation. Since the radiative emission is nearly collinear, the hard $(x \sim 1)$ and intermediate $(1 \gg x \gg T/E)$ fragments largely remain confined within a narrow angular cone $(\theta \leq 0.16)$. Even at the later stages of the evolution ($t=4.1,9.8 {\rm fm}/c$) when the hard ($x\sim 1$) peak has significantly depleted, the hard and intermediate partons largely remain within such a narrow cone, except for rare large angle scatterings that will be discussed further below.

Energy lost by the hard fragments accumulates in the soft sector starting at momentum fractions $x \lesssim 4\pi T/E$, where the energy distribution shows a pronounced peak. Since elastic scattering are most efficient at these soft scales, they lead to sizable broadening out to cone-sizes $R \gtrsim 0.3$ already at early times.  Over the course of the evolution, broadening of the soft sector continues to drive energy out of the collinear region to larger angles. By the time $t=9.8 {\rm fm}/c$, most of the energy resides in the soft sector ($x \lesssim 4\pi T/E$) and has been transported to large angles between $R=0.3$ and $R=0.6$ cone size. Eventually, once the residual hard fragments have lost all of their energy, the evolution reaches the asymptotic distribution in Eq.~(\ref{eq:DeqSolMain}), whose angular structure will be further discussed below.

\subsection{Angular structure of the cascade}
\begin{figure}
    \centering
    \includegraphics[width=0.98\textwidth]{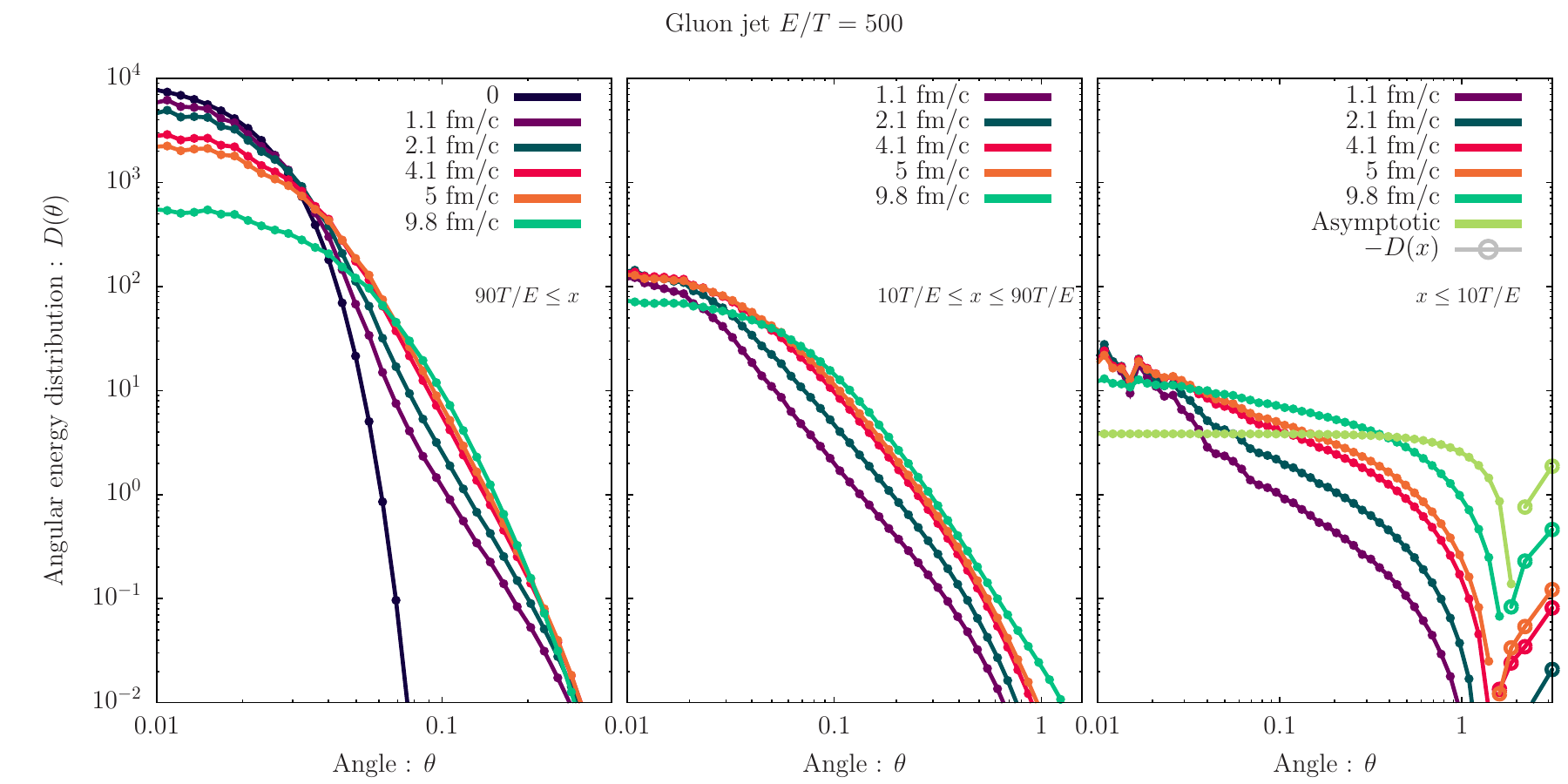}
    \caption{Evolution of the angular distribution in different momentum regions, as described by Eq.~(\ref{eq:AngularDist}).}
    \label{fig:AngularDist}
\end{figure}
We will now investigate the angular structure of the energy cascade and study the redistribution of energy out to large angles. Our results are compactly summarized in Fig.~\ref{fig:AngularDist} where we analyze the angular structure of the energy distribution by dividing the spectrum into three energy bins corresponding to the large momentum sector ($xE \geq 90~ T$) in the left panel, intermediate scales ($ 10~ T \leq xE \leq 90~T $) in the middle and the soft sector ($ xE \leq 10~T $) in the right. Each panel shows the integrated energy distribution as follows
\begin{equation}\label{eq:AngularDist}
    D(\theta,t)|_{x_{\rm min}}^{x_{\rm max}} = \sum_{a} \int_{x_{\rm min}}^{x_{\rm max}} \rmd x~ D_{a/\rm jet}(x,\theta,t)\;.
\end{equation}
Starting with the hard sector, we observe that the peak is only marginally broadened as it is depleted from early to late times and energy is transferred to intermediate and lower momentum scales $(x\leq 90T/E)$. Similar effects are seen at intermediate scales, where the distribution displays a rapid fall off at large angles while the peak is depleted from the collinear region. Due to the constant influx of energy at small angles $\theta \ll 1$, the energy distribution in the soft sector also remains peaked at small angles; nevertheless, over the course of the evolution more and more of the energy in the soft sector is redistributed out to large angles $\theta\sim 1$ due to elastic interactions. One also observes, that throughout the entire evolution the energy distribution in the soft sector features a negative away side peak, which accounts for the depletion of thermal QGP particles moving in the opposite direction of the jet that is necessary to account for momentum conservation.

Summarizing the results in Fig.~\ref{fig:AngularDist}, we observe that for typical cone-sizes $\theta \sim 0.3$ the angular broadening at high and intermediate scales is a sub-leading effect. Instead, the re-distribution of energy to large angles proceeds predominantly via the collinear radiative break-up of the hard parton, with subsequent broadening of the soft sector where the elastic processes are more efficient in redistributing energy out to large angles.

\begin{figure}
    \centering
    \includegraphics[width=0.6\textwidth]{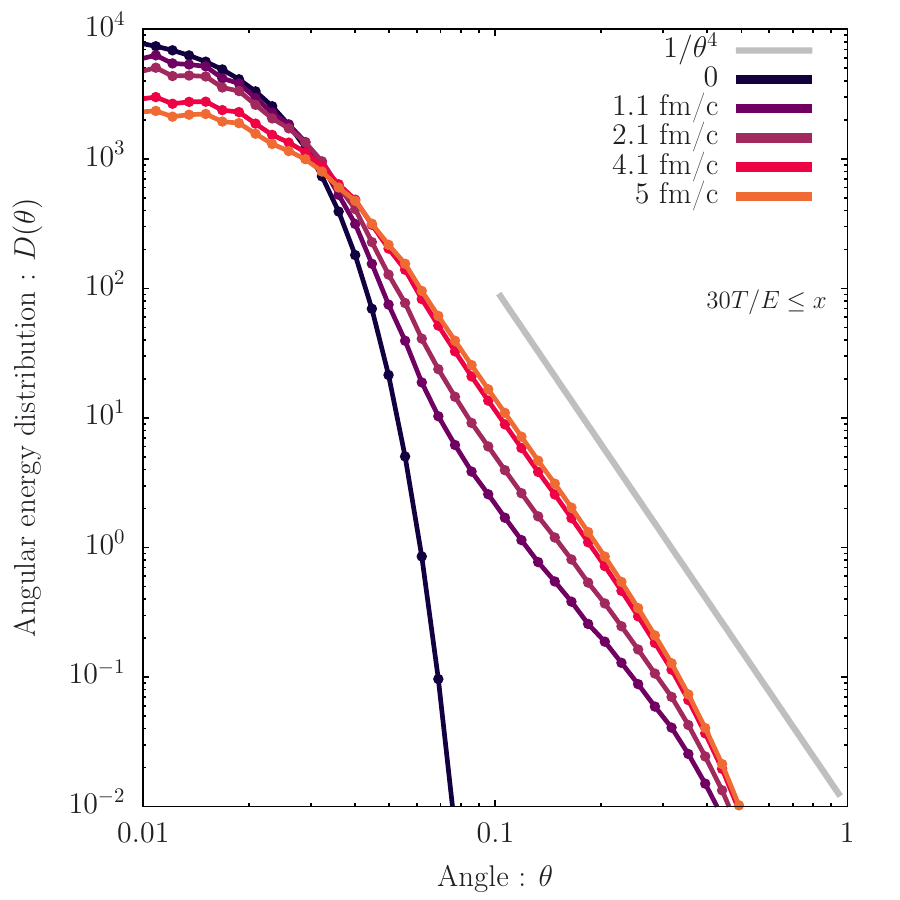}
    \caption{Evolution of the angular distribution for hard particles with momentum fraction $x>30~T/E$. The gray line that represents the expected Coulomb tail $1/\theta^4$ at large angles is shown for reference.
    }
    \label{fig:AngularTail}
\end{figure}

Despite the fact that the broadening of high momentum particles provides a subleading effect for re-distribution of energy to large angles, it is interesting to investigate this phenomenon more closely, as is done in Fig.~\ref{fig:AngularTail}, where we show the angular distribution $D(\theta,t)$ of high-energy fragments $x>30T/E$ at various times of the evolution. Starting from early times where the distribution is extremely narrow, one observes a simultaneous broadening and depletion of the peak along with the emergence of a power-law tail at intermediate angles, which can be attributed to rare large angle scatterings~\cite{Kurkela:2014tla} -- as originally discussed by Moli\`ere theory for QED \cite{Moliere:1947zz}. Specifically for high energy fragments $x \gg T/E$, the effects of elastic collisions on the evolution of the energy distribution $D_{a/\rm jet}(x,{\bf q}_\bot,t)$ as a function of energy fraction $x$ and transverse momentum ${\bf q}_\bot$ 
\begin{align}
\label{eq:FragmentationFctMom}
D_{a/\rm jet}(x,{\bf q}_\bot,t)=\nu_a\int d^3\mathbf{x} \int \norm{p}~\frac{|\p|}{E}~\delta\Big(\frac{|\p|}{E}-x\Big)\delta^{(2)}\Big({\bf q}_\bot -{ \bf p}_{\bot}\Big) \delta \bar{f}_a(\p,\mathbf{x},t)\;,
\end{align}
can be approximately described by~\cite{Blaizot:2014rla,Mehtar-Tani:2018zba,Schlichting:2020lef}
\begin{align}
\label{eq:simplebranchingandbroadening}
&\partial_{t}D_{a/jet}(x,{\bf p}_\bot,t) \simeq \int_{{\bf q}_{\bot}} C_{a}({\bf q}_\bot) \left[ D_{a/jet}(x,{\bf p}_\bot-{\bf q}_\bot,t) -D_{a/jet}(x,{\bf p}_\bot,t) \right] \nn
& +\sum_{bc} \int_{0}^{1}dz~\left[ \frac{1}{z^2}\frac{d\Gamma^{b}_{ac}}{dz}\Big(\frac{xE}{z},z\Big) D_{b/jet}\Big(\frac{x}{z},\frac{{\bf p}_{\bot}}{z},t\Big)- \frac{1}{2} \frac{d\Gamma^{a}_{bc}}{dz}\Big(xE,z\Big) D_{a/jet}\Big(x,{\bf p}_{\bot},t\Big) \right]\;,
\end{align}
where as in Eq.~(\ref{eq:simplebranchingandbroadening}), the first line describes the effects of radiative emissions. The second line in Eq.~(\ref{eq:simplebranchingandbroadening}) describes transverse momentum broadening due to elastic interactions (while neglecting long. momentum and energy transfer as well as species changing processes), as described by the collisional broadening kernel $C_{a}({\bf q}_\bot)$, which has been determined in \cite{Arnold:2008vd,Aurenche:2002pd} for leading order QCD kinetic theory. Within this framework, the rare large angle scatterings correspond to a $C_{a}({\bf q}_\bot) \sim 1/{\bf q}_{\bot}^4$ power-law tail of the collisional broadening kernel, for momentum transfers $|{\bf q}_\bot| \gg m_{D}$. When expressed in terms of angular variables, the momentum transfer is given by $|{\bf q}_\bot| = xE~\sin\theta \simeq xE~\theta$, such that the angular distribution $D(x,\theta)$ can be expected to follow a  $\sim 1/\theta^4$ behavior at intermediate angles where large angle scatterings dominate. While small angle scatterings lead to a diffusive broadening at small angles, we observe in Fig.~\ref{fig:AngularTail} that the $\sim 1/\theta^4$ power law indicated by the gray dashed line dominates the large angle region up to angles $\theta \sim 0.4$, beyond which the energy distribution drops sharply as energy and momentum conservation further restrict the available phase space for large angle scatterings.

\section{Sensitivity to soft fragments}\label{sec:Soft}

Based on our analysis, we noted that only the soft fragments of the medium induced shower escape to large angles, while partons with momenta larger than the temperature stay mostly collinear throughout the evolution. In this section, we will further explore the consequences of this behavior by comparing the energy carried by soft and hard fragments inside the cone. This question is of both conceptual and phenomenological relevance. Conceptually, in the standard approach to jet quenching the degradation of energy due to the radiative break-up cascade can be discussed without an explicit microscopic description of the dissipation of energy in the thermal medium. This is due to the so-called inviscid property of turbulence that states that the turbulent cascade transports energy from high to low momentum scales, regardless of the detailed properties of the hard source and the dissipation of energy in the thermal medium. We recall that the effect of dissipation is to stop the turbulent cascade by causing the accumulation of energy around scales $\sim 2\pi T$ and its subsequent thermalization. Hence, when a proper microscopic description of the dynamics of the soft sector is included, the turbulent cascade may form only well  above the temperature scale. 

We may therefore wonder to what extent the details of the kinetic thermalization process and the resulting deformation of the spectrum affect energy loss, as compared to a purely high-momentum description that ignores detailed balance at scales $\lesssim 2\pi T$. This question is crucial as far as the predictive power of the theory is concerned. While the QCD coupling constant can be treated as small for the radiative break-up of high-energy fragments, such that perturbation theory can be justified and systemically improved, the dynamics near the temperature scale may in general turn out to be non-perturbative in which case our kinetic description can at best be viewed as a perturbative model of the soft sector. 

Beyond the conceptual importance of clarifying the role of low scale  thermalization in jet quenching, this question is also of considerable phenomenological importance. In particular, it would be important to quantify the sensitivity of jet observables to the dynamics of soft thermalization, in order to devise jet observables that can be accurately calculated within a perturbative QCD approach on the one hand 
and on the other hand explore means to study QCD thermalization via jet quenching in heavy-ion collisions. 

While our kinetic framework is in principle well suited to provide an understanding of the energy recovery due to medium response and recoil, which constitutes a highly active area of research in QCD be quenching~\cite{Casalderrey-Solana:2004fdk,Renk:2006mv,Qin:2009uh,Li:2010ts,Tachibana:2014lja,Tachibana:2017syd,Casalderrey-Solana:2020rsj,Yang:2022nei}, a detailed phenomenological analysis is well beyond the scope of the present work. Nevertheless, we will attempt to provide some new insights into this question by comparing the in-cone energy with and without the soft modes that fall below $2\pi T$, which can be viewed as an {\it ad hoc } energy cutoff of the radiative break-up cascade. 

\begin{figure}
    \centering
    \includegraphics[width=0.5\textwidth]{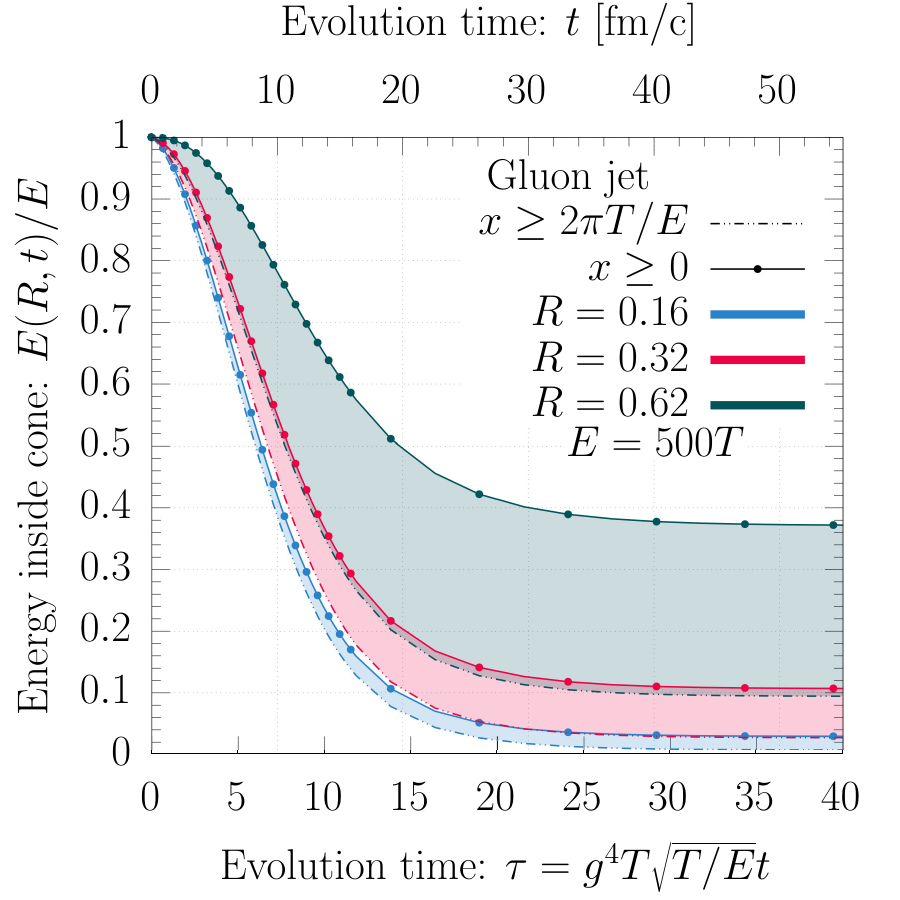}\includegraphics[width=0.5\textwidth]{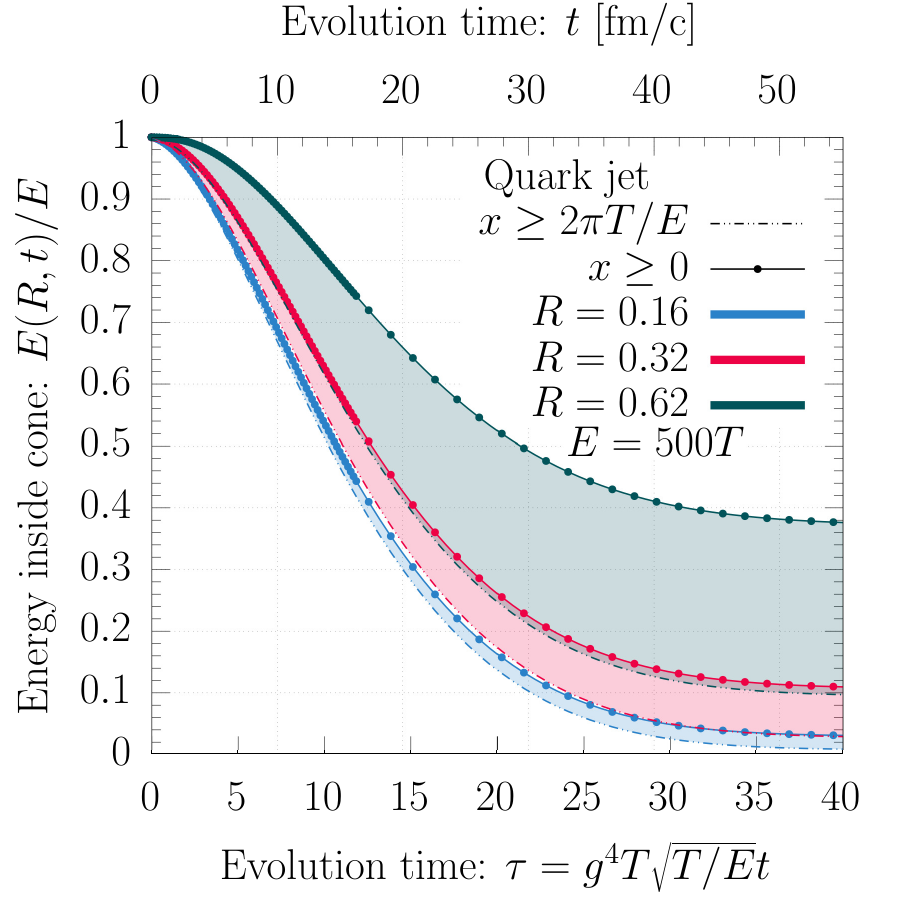}
    \caption{Evolution of the energy inside the cone ($\theta\leq R$) for gluon (left) and quark (right) initial jets with $E=100T$. Full line-points represent the full momentum region as in Eq.~(\ref{eq:FullCone}), while the dashed lines describe only the high momentum fraction region $xE\geq 2\pi T$ as in Eq.~(\ref{eq:HighCone}).}
    \label{fig:AngularJetEnergyLoss}
\end{figure}

The energy remaining inside the jet cone of size $R$ is computed by integrating the energy distribution up to an angle $R$
\begin{equation}\label{eq:FullCone}
    E(R,\tau) = E\sum_{a} \int \rmd x~\int_{\cos R}^{1}\rmd\co~ D_{a/\rm jet}(x,\theta,\tau)\;.
\end{equation}
Now, for the energy carried by hard partons in the cascade only the integration over the momentum fraction $x$ is cutoff at the low momentum $(x\geq 2\pi T/E)$
\begin{equation}\label{eq:HighCone}
    E_{\rm hard}(R,\tau) = E\sum_{a}\int_{2\pi T/E}^{\infty} \rmd x~\int_{\cos R}^{1} \rmd\co~ D_{a/\rm jet}(x,\theta,\tau)\;.
\end{equation}
Our results are shown in Fig.~\ref{fig:AngularJetEnergyLoss} for three cone-sizes $R=(0.16,0.32,0.62)$ (green, red, blue) for an initial gluon jet in the left panel and an initial quark jet in the right panel. Solid lines with points at the upper edge of the band represent the full energy inside the cone in Eq.~(\ref{eq:FullCone}), while the dashed lines at the lower edge of the band represent only the contribution from the hard momentum region in Eq.~(\ref{eq:HighCone}), such that the bands showcase the discrepancy between the two. 

We find that for narrow cones $(R\leq 0.3)$ the dynamics of the soft sector does not play a major role as the energy loss is very similar in both momentum regions.
However, for larger cone sizes, a transition occurs where the initial energy loss becomes slower, and we observe a significant discrepancy between the two curves, which highlights the sensitivity to the dynamics of the soft sector for cone sizes $R\gtrsim 0.3$. Clearly, the physical origin of this discrepancy can be easily understood, as the energy loss in the hard sector is driven by the depletion due to the in-medium splitting, while the energy loss in the soft sector is due to elastic scatterings which re-distribute the energy to large angles outside the cone. While frequent small angle elastic scatterings induce a sufficient amount of broadening for narrow cone sizes, with increasing cone size the energy transport to large out-of-cone angles takes an increasing amount of time. We can therefore anticipate, that for larger cone sizes, the out-of-cone energy loss shows a larger sensitivity to the detailed dynamics of the thermalization of the soft sector, which is confirmed by our analysis in App.~\ref{ap:Small-Angle} where we compare numerical results with full HTL elastic scattering to the corresponding ones obtained in the small-angle/diffusion approximation~\cite{Schlichting:2020lef}.

Eventually  at late times, where the jet has nearly equilibrated, the energy inside the cone slowly approaches a constant value, which depends only on the cone size $R$. Indeed, the amount of energy that remains inside the jet cone can be estimated from the asymptotic distribution as
\begin{align}
    E_{\rm eq}(R) =& E \sum_{a} \int \rmd x~\int_{\cos R}^{1} \rmd \co~ D_{a}^{\rm eq}(x,\theta)
    = E \left[ \frac{5}{2} + \frac{3}{2} \cos (R)\right] \sin^2 \Big( \frac{R}{2} \Big)\;, \label{eq:EnergyDistributedEQ}
\end{align}
such that for small cone sizes $R \ll 1$ one has $E_{\rm eq}(R) \sim E R^2$. While Eq.~(\ref{eq:EnergyDistributedEQ}) provides the cone size dependence of the asymptotic energy distribution, we observe from Fig.~\ref{fig:AngularJetEnergyLoss} that irrespective of the cone size, the soft fragments carry a substantial fraction of this energy $(\gtrsim 60\%)$, at late times when the jet has nearly equilibrated.

\begin{figure}
    \centering
    \includegraphics[width=\textwidth]{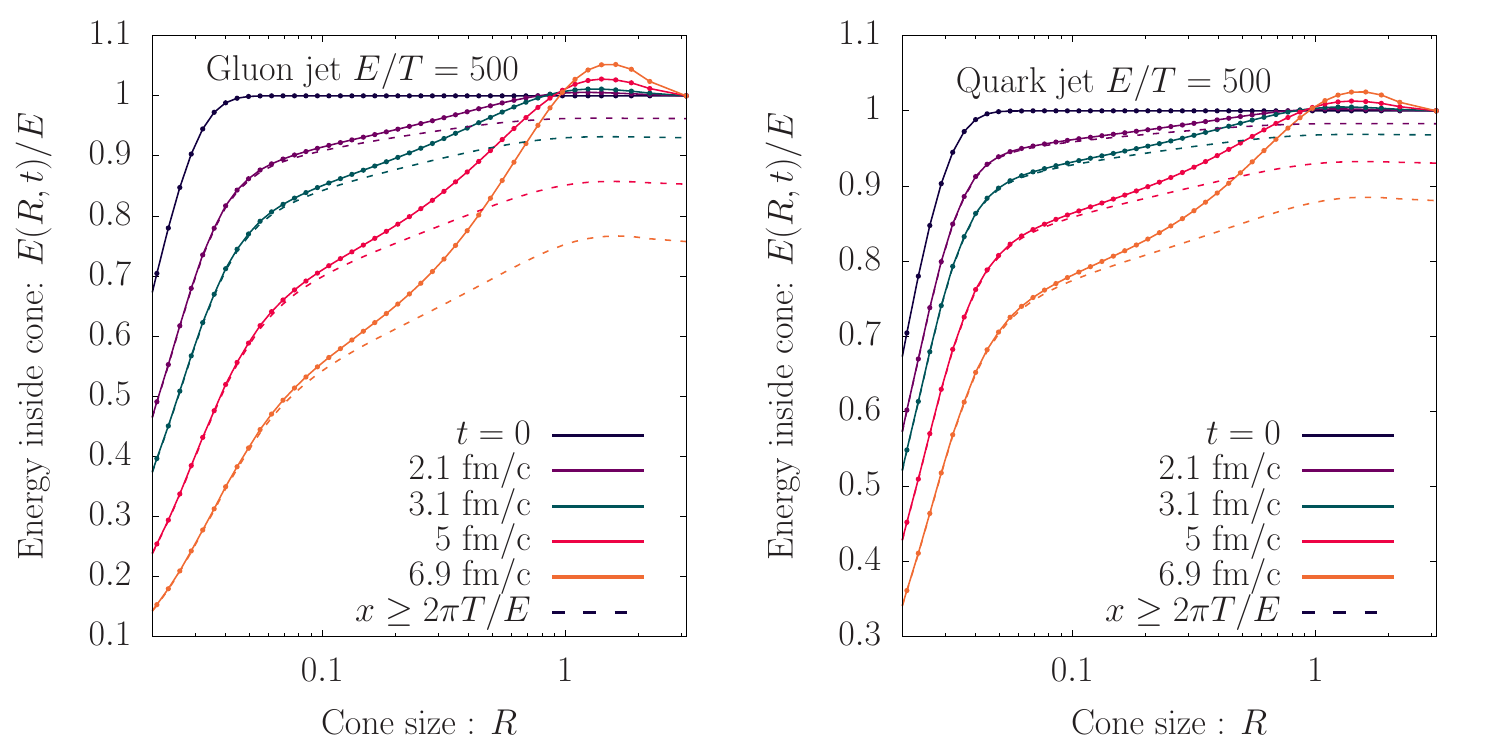}
    \caption{Evolution of the energy stored inside the cone $R$ as a function of the cone size $R$. (left) Gluon jet, (right) Quark jet. Linespoints represent the energy carried by all partons while dashed lines represent the energy carried by hard partons with momentum fraction $x\geq 2\pi T/E$.}
    \label{fig:AngularEnergy}
\end{figure}
Next, in order to showcase the relevant angular scale where the separation between soft and  hard partons is relevant for the out-of-cone energy loss, we display in Fig.~\ref{fig:AngularEnergy}, the cumulative energy $E(R,\tau)$ inside a cone as a function of the cone size $R$ for different times $t=(2.1,3.1,5,6.9)$fm/c with an initial gluon jet in the left panel and an initial quark jet in the right panel. Solid lines with points represent the full energy inside the cone, while dashed lines represent the energy carried by hard partons with momentum fraction $x\geq 2\pi T/E$ only. Note that the angular scale is logarithmic. Starting from early times, one observes a strong depletion of the energy inside narrow cones, along with a slight broadening of the core structure, which is better visible for the gluon jet that experiences more broadening. Eventually, as the energy is depleted from the initial hard parton at $x\sim 1$ and $R\sim 0$, we observe the emergence of an angular scale, where the energy inside narrower cones is mostly carried by hard partons while for larger cone sizes the soft sector gradually starts to contribute a substantial fraction of the energy. Over the course of the evolution, this scale evolves from $R \sim 0.03$ at $t=2.1 {\rm fm}/c$ to around $R \sim 0.1 $ at $t=6.9 {\rm fm}/c$. Since the contribution of hard fragments saturates at large angles, we can also confirm that the hard partons mostly stay collinear throughout the evolution, and except for very narrow cones, the out-of-cone energy loss occurs primarily due to the re-distribution of soft fragments out to large angles. While the soft sector always dominates the energy distribution at large angles, the precise scale where this transition happens depends on the evolution time, i.e. on the amount of quenching that the jet has suffered inside the medium. Beyond cone size of $R\sim 1$, the distribution overshoots unity and decreases again due to the negative  contributions of recoiling partons, before eventually all energy is recovered at $\theta=\pi$ where $E(\pi,\tau)=E$ at all times.

\section{Stopping time scaling of energy loss }\label{sec:Stopping}
\begin{figure}
    \centering
    \includegraphics[width=\textwidth]{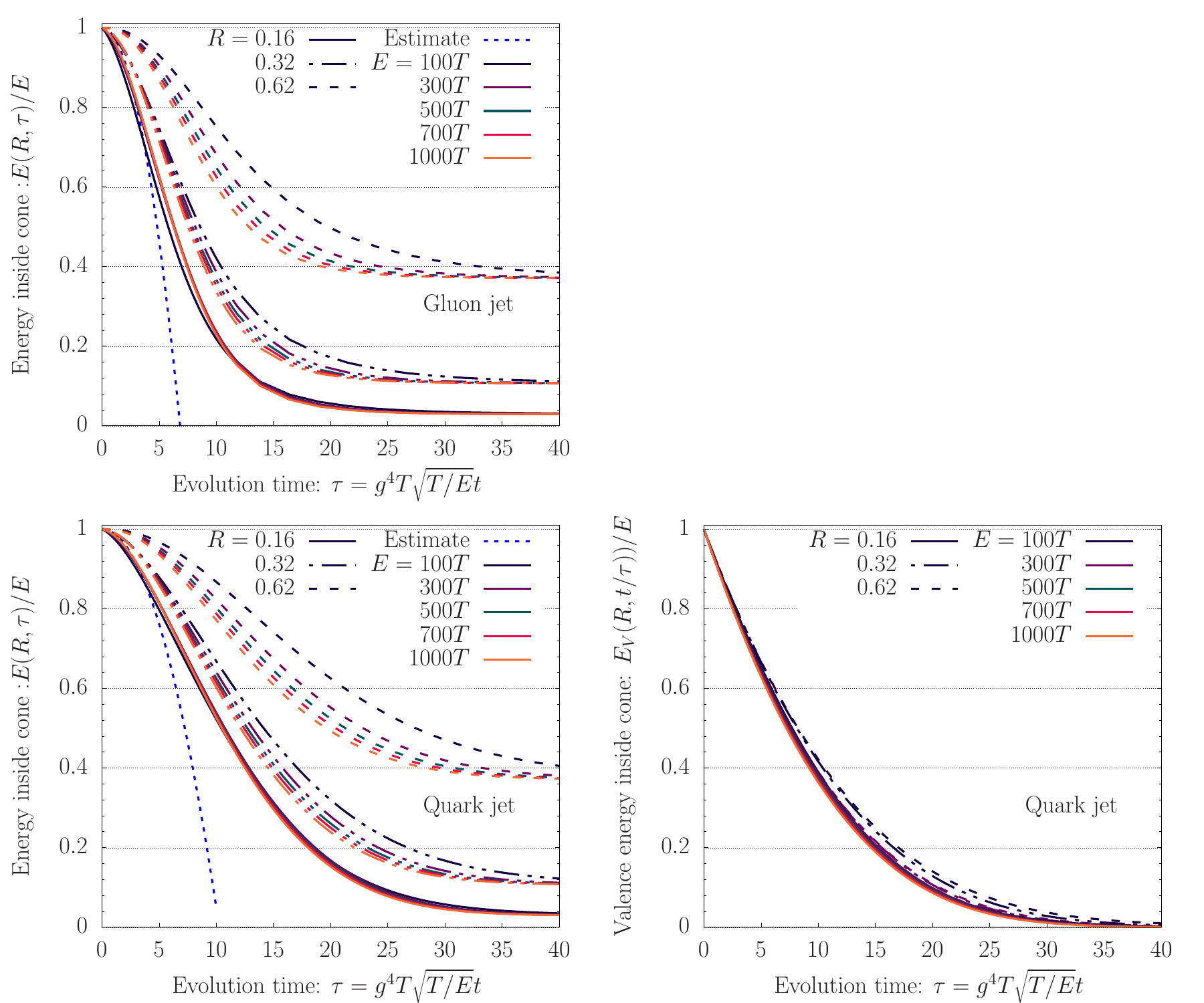}
    \caption{Evolution of the fraction of the initial energy carried by particles inside cones $R=0.16,0.32,0.62$ with different initial energies $E=100-1000T$. The left panels represent the evolution of the total energy $E(R,\tau)/E=\sum_a E_a/E$, starting with a gluon initial jet (top) or quark initial jet (bottom). The bottom right panel represents the valence energy, defined as the difference of energy carried by quarks and anti-quarks $E_{V}(R,\tau)/E=(E_q - E_{\bar{q}})/E$.}
    \label{fig:DifferentEnergies}
\end{figure}
So far, we have analyzed the in-medium evolution for a fixed initial energy $E=500T$ of the hard parton. Based on our observation that the out-of-cone energy loss proceeds via radiative break-up of hard fragments, followed by angular broadening of soft fragments, we can anticipate that the evolution of the out-of-cone energy loss is governed by the splitting time $t_{\rm th}\sim \frac{1}{\alpha_s} \sqrt{\frac{E}{\hat{q}}}$ of the initial hard parton in Eq.~(\ref{eq:SplittingTime}) that governs the energy deposition into the soft sector. Verifying to what extent a simple scaling with $t_{\rm th}$ holds in our numerical simulations is thus  not only important for phenomenological applications, but also provides an important verification of the underlying energy loss mechanism. 

Our results for the energy dependence of the out-of-cone energy loss are compactly summarized in Fig.~\ref{fig:DifferentEnergies}, where we display the energy remaining inside a cone of size $R=0.16,0.32,0.62$. By scaling the evolution time in terms of the typical splitting time $t_{\rm th}$ as in Eq.~(\ref{eq:TauScaling}), we account for the leading energy dependence when comparing different initial parton energies $E=(100,300,500,700,1000)T$ for an initial gluon jet in the top panel and initial quark jet in the bottom left panel. 

We also compare the evolution with the analytical estimates computed for the energy deposition in the soft sector due to the turbulent cascade, which leads to the following linear energy loss rate  \cite{Mehtar-Tani:2018zba,Schlichting:2020lef} 
\begin{align}
    \frac{1}{E}\frac{dE}{d\tau} = \frac{C_A^{1/2}C_{R}}{8\pi^2} \sqrt{\frac{\hat{\bar{q}}(E)}{g^4T^3}} \left(\tilde\gamma_g  +\frac{S}{G} \tilde\gamma_q\right)  \tau\;,\label{eq:EnergyLossRate}
\end{align}
where $C_{R}=C_{A}$ for gluon and $C_{R}=C_{F}$ for quarks jets, $S/G=0.07\times 2 N_f$ is the quark to gluon ratio in the turbulent cascade \cite{Schlichting:2020lef}, and 
\begin{align}
    \label{eq:TildGammag}
     \tilde\gamma_g =& \int_{0}^1 \rmd z ~  \sqrt{\frac{xE}{g^{8}T^3}}\Big[\frac{d\Gamma^{g}_{gg}}{dz}\Big(xE,z\Big) +2N_f
    \frac{d\Gamma^{g}_{q\bar{q}}}{dz}\Big(xE,z\Big) 
     \Big]~z\log(z)\nn
    =& \frac{1}{8\pi^2}\sqrt{\frac{\hat{\bar q}(\sqrt{TE})}{g^4T^3}}(25.78 +2N_f0.177)\;, \\
    \label{eq:TildGammaq}
    \tilde\gamma_q =& \int_{0}^1 \rmd z ~  \sqrt{\frac{xE}{g^{8}T^3}}\Big[\frac{d\Gamma^{q}_{gq}}{dz}\Big(xE,z\Big)+\frac{d\Gamma^{q}_{qg}}{dz}\Big(xE,z\Big)\Big]~2z\log(z) 
    =  \frac{1}{8\pi^2}\sqrt{\frac{\hat{\bar q}(\sqrt{TE})}{g^4T^3}}(11.595)\;, 
\end{align}
are flux constants governing the energy flux along the turbulent cascade, as discussed in detail in~\cite{Schlichting:2020lef}.\footnote{Note that to evaluate the integrals, one employs the leading logarithmic approximation of the splitting rates~\cite{Arnold:2008zu} where $\sqrt{\frac{xE}{g^{8}T^3}} \frac{d\Gamma^{a}_{bc}}{dz}\Big(xE,z\Big)$  becomes approximately independent of the coupling constants $g$ and the energy $xE$ of the emitter, e.g.
$$\sqrt{\frac{xE}{g^{8}T^3}} \frac{d\Gamma^{g}_{gg}}{dz}\Big(xE,z\Big) \simeq \frac{1}{8\pi^2} P_{gg}(z) \sqrt{\frac{\hat{\bar{q}}(xE)}{g^{4} T^3}}~\sqrt{\frac{(1-z)C_A + z^2 C_A }{z(1-z)}}\;.$$
We evaluate this at the scale $xE=\sqrt{TE}$ which is the geometric mean of the hard $E$ and thermal scale $x=T$. In practice, when we present our numerical estimate we extract $\sqrt{\frac{\hat{\bar{q}}(E)}{g^4T^3}} \overset{E=1000T}{\simeq}1.39$ and
$\sqrt{\frac{\hat{\bar{q}}(\sqrt{TE})}{g^4T^3}} \overset{E=\sqrt{1000}T}{\simeq} 1.20$ from fits to the numerical results for the in-medium splitting rates, as discussed in App.~A of \cite{Schlichting:2020lef}. Collecting everything this yields $E(\tau)= E\big(1-  0.014 C_{R} C_{R} \frac{\tau^2}{2}\big)$ as the estimate for the energy loss, which is depicted in Fig.~\ref{fig:DifferentEnergies}.
} 
By assuming that the energy in the soft sector escapes the jet cone on a time scale that is much smaller than the energy deposition due to the turbulent cascade, the linear energy loss rate in Eq.~(\ref{eq:EnergyLossRate}), then leads to a quadratic decrease of the energy inside the cone, which is shown as a dashed line in Fig.~\ref{fig:DifferentEnergies}. 

When considering the energy inside a cone, we find that the scaling is more effective for narrow cone sizes $R\leq 0.16$, where the energy deposited in the soft sector escapes the cone quasi-instantaneously on the time scale of the energy loss of the hard partons. In this regime, the out-of-cone energy loss is also remarkably well described using the analytical estimate in Eq.~(\ref{eq:EnergyLossRate}). Conversely, for larger cone sizes the scaling only holds approximately and gradually starts to break down as the time scale for the re-distribution of energy to large angles becomes significant, leading to a breakdown of the simple scaling with the time scale for in-medium splittings. Energy loss to large angles is then also sensitive to the thermalization of the soft sector, and can no longer be captured by simple estimates as in Eq.~(\ref{eq:EnergyLossRate}).

We also show the valence energy by taking the difference of the energy carried by quarks and anti-quarks in the bottom right panel. Since, the energy difference of quarks and anti-quarks is not linked to a conserved quantity, there is an actual loss of energy in the valence channel, primarily due to in-medium splittings of hard valence quarks. We observe that the valence energy exhibits virtually no dependence on the cone-size, as the momentum broadening of the leading valence parton is negligible. Since the energy loss in the valence channel is governed by in-medium splittings, the results for different initial energies scale with the splitting time to very good accuracy.

\begin{figure}
    \centering
    \includegraphics[width=0.5\textwidth]{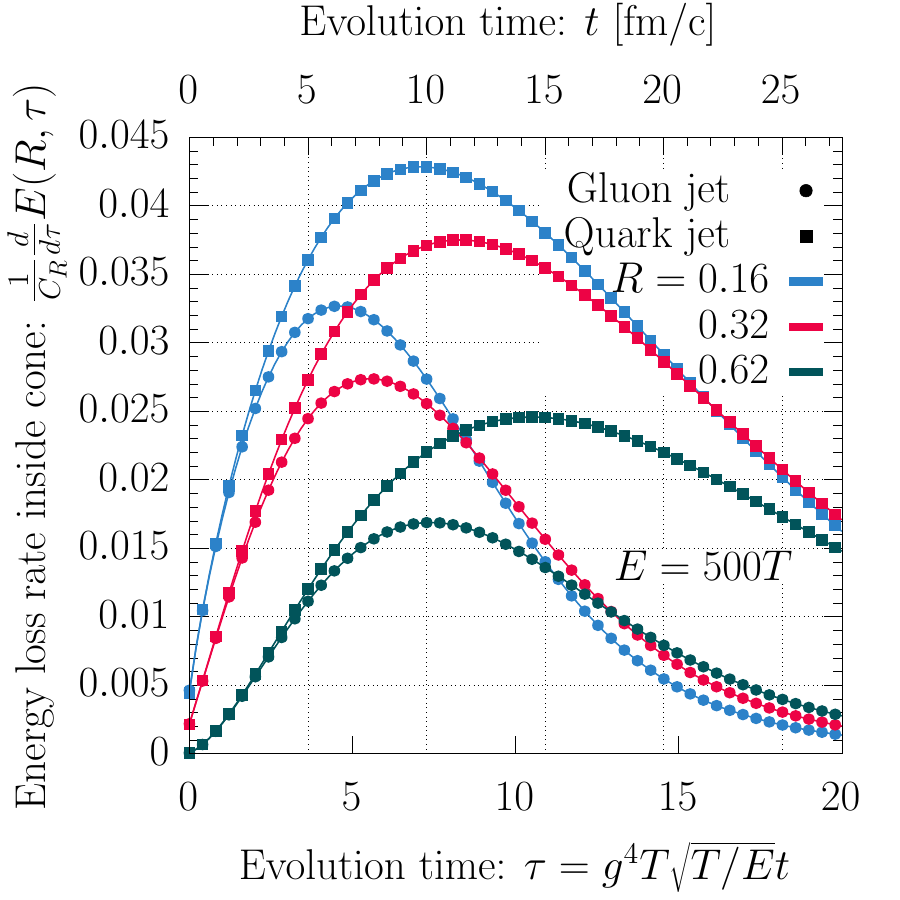}\includegraphics[width=0.5\textwidth]{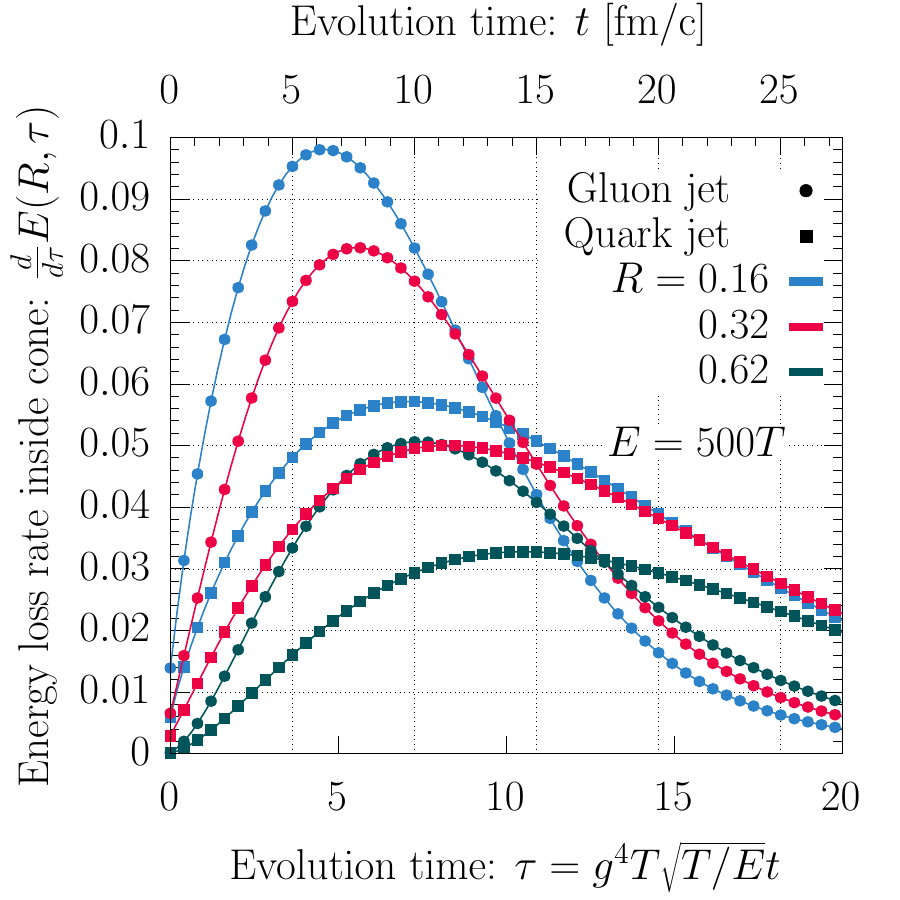}
    \caption{Evolution of the energy loss rate for different cone sizes $R=0.16,0.32,0.62$, for a Gluon (square) and a Quark (circle) jet.
    The left panel shows the rate divided by the Casimir of the inital jet, displaying the linear behavior at early times. }
    \label{fig:EnergyLossRate}
\end{figure}
We investigate the scaling of the energy loss rate between the Quark and Gluon jets.
By performing a time derivative of the energy remaining inside the cone of size $E(R,\tau)$ from Eq.~(\ref{eq:FullCone}),
we obtain the energy loss rate $\frac{dE(R)}{d\tau}$ shown in Fig.~\ref{fig:EnergyLossRate}.
The energy loss rate is linear at early times for small to intermediate cone sizes $R=0.1 - 0.3$, while for larger cone sizes it displays a non-linear behavior.
For all cone-sizes we observe a clear Casimir scaling with the initial jet, which governs both elastic interactions and in-medium splittings.
However, the Casimir scaling breaks down at late time when the flavor content of the cascade becomes more complex than the simple initiating parton \cite{Mehtar-Tani:2018zba,Schlichting:2021idr}.

\section{Modeling medium quenching}\label{sec:Quench}
\begin{figure}
    \centering
    \includegraphics[width=0.5\textwidth]{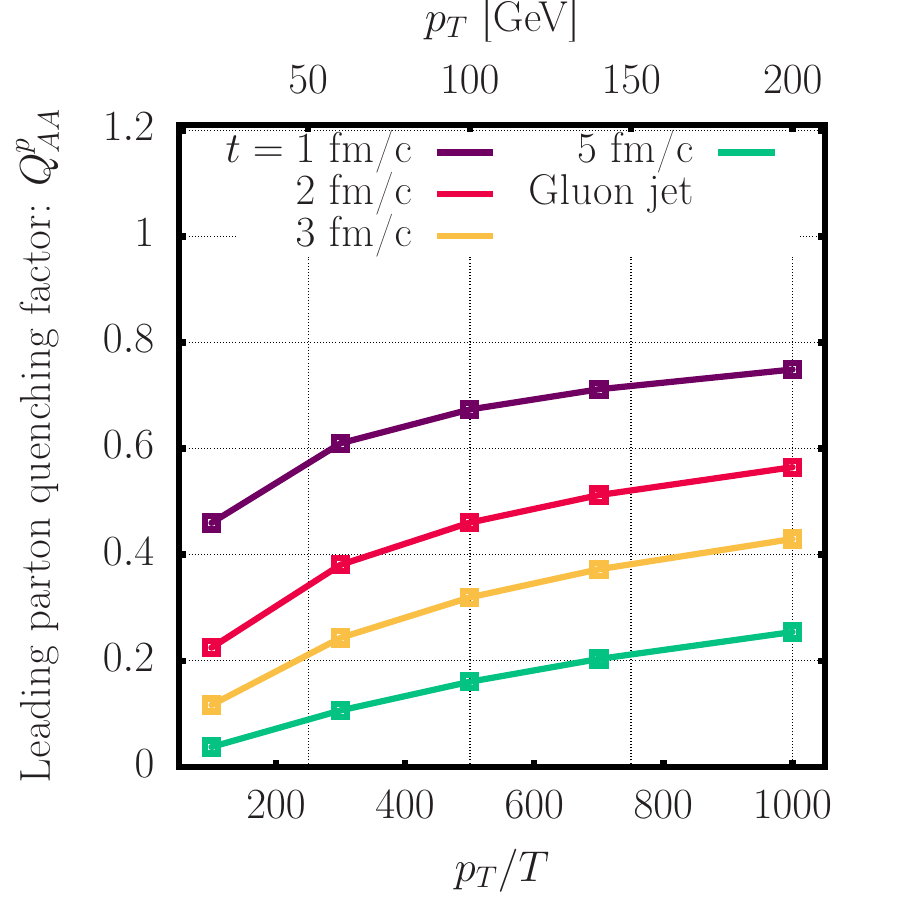}
    \includegraphics[width=0.5\textwidth]{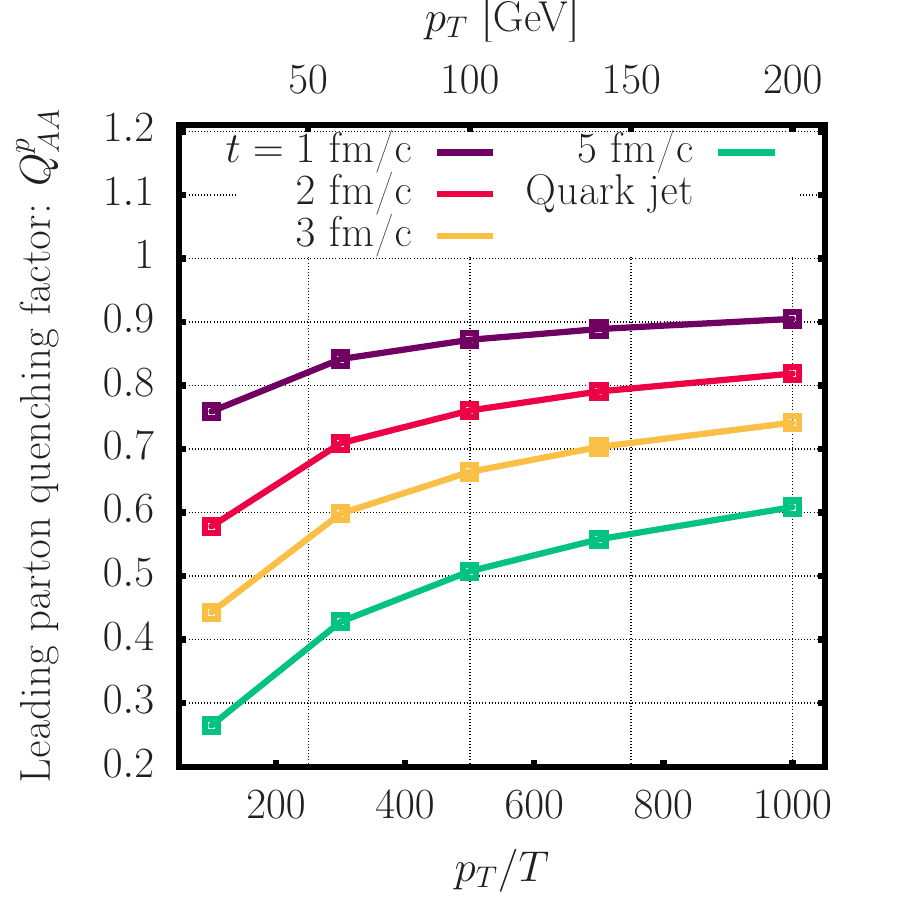}
    \caption{(left) Leading parton quenching factor as a function of initial parton energy. We take the medium temperature $T=200$MeV. }
    \label{fig:LeadingParton}
\end{figure}
Our investigations of energy loss in the previous sections have been of rather theoretical nature, and we would now like to illustrate how our results can be used in phenomenological studies of jets in heavy-ion collisions. Such studies rely on the assumption that the hard partonic process that initiates the jet is independent from the subsequent final state interactions with the QGP.  This is rooted in the principle of factorization of short distance/time physics, that take place over scales of the order $1/p_T$, and thus is independent of the QGP dynamics so long as the in-medium mean free path is larger than $1/p_T$, and long distance physics where the interactions of the jet with the plasma constituents take place, which lead to jet energy loss and its substructure modifications \cite{Mehtar-Tani:2017web,Qiu:2019sfj}. By comparing with proton-proton collisions where no final state interactions occur, one can infer quantitatively the effects of medium modifications. 

Since the results of our kinetic evolution can be seen as a Green's function for the evolution of a (resolved) hard parton inside the medium, we can compute the suppression of hard partons by folding our results of energy loss with hard partonic spectrum from proton-proton collisions. However, our effective kinetic evolution lacks important physics due to vacuum like emissions and path length dependence of the in-medium radiation rates, as well as a proper space-time evolution of the medium.  
While these aspects are known to be important for phenomenological studies, we will nevertheless employ our results to obtain estimates for quenching of spectra yields in order to illustrate their use and gain qualitative insights.

\subsection{Leading parton quenching}
Firstly, we consider the yield of the inclusive parton spectrum after passing through the medium which can be obtained as a convolution of the energy distribution for a given starting energy $p^{in}_T$ with the initial parton spectrum\footnote{We assume parton-hadron duality, i.e., the fragmentation does not substantially alter the leading parton such that each leading parton leads to a high-$p_T$ hadron in the final state.} \cite{Baier:2001yt,Salgado:2003gb,Mehtar-Tani:2014yea,Adhya:2019qse}
\begin{align}
    \frac{d^2\sigma_{AA}}{dp^2_T}(p_T) =& \int_0^{\infty} \rmd^2p^{in}_T  \int_0^1 \frac{\rmd x}{x}\int_{-1}^{1} \rmd\co ~ \delta^{(2)}(p_T -xp^{in}_T) \nonumber\\
    &D\left(x,\theta,\tau\equiv g^4 T \sqrt{T/p_T^{in}} t \right)\frac{\rmd^2\sigma_{\rm vac}}{\rmd^2p^{in}_T}(p^{in}_T)\;,
\end{align}
where we make use of the rescaled time $\tau = g^4T\sqrt{\frac{T}{E}} t$ from Eq.~(\ref{eq:TauScaling}) \cite{Adhya:2019qse} in order to account for the initial parton energy (c.f. Fig.~\ref{fig:DifferentEnergies} for the energy dependence). Dividing by the vacuum spectrum approximated by a power-law $\frac{d^2\sigma_{\rm vac}}{dp^2_T}(p^{in}_T)\propto p_T^{-1-n}$, we define the inclusive leading parton quenching factor
\begin{align} \label{eq:Q-int}
    Q^{\rm p}_{AA}(p_T) =& \frac{\frac{d^2\sigma_{AA}}{\rmd p^2_T}}{\frac{\rmd^2\sigma_{\rm vac}}{\rmd p^2_T}}
    = \int_0^{1} \rmd x\int_{-1}^{1} \rmd\cos\theta~  D\left(x,\theta,g^4 T \sqrt{xT/p_T} t\right) x^{n-2}\;.
\end{align}

We follow \cite{Spousta:2015fca} and take $n_g=5.66$ and $n_q=4.19$ for gluon and quark jets, respectively. The resulting leading parton quenching factors are shown in Fig.~\ref{fig:LeadingParton}, for different times $t=1,2,3,5$fm/c and for both gluon and quark jets.
We will not directly compare with experimental data, still we observe a similar qualitative behavior to the suppression $R_{AA}$ of charged hadrons \cite{STAR:2002ggv,PHENIX:2005zfm,CMS:2012aa,ATLAS:2018bvp,STAR:2016jdz,PHENIX:2020alr}. The quenching factor display a suppression ($Q<1$) in the full range of energy $p_T =100-1000T$, where higher energies are less suppressed as expected. However, since we treat medium-induced splitting in the infinite medium limit which tends to overestimate the rates, the quenching factor we obtain at comparable times of heavy-ion collisions ($t\sim 5$fm/c) is much more suppressed than experimental results. Hence, we conclude that other effects such as the path length dependence of radiative emission rates are important to reproduce experimental observables. 
As expected, we find that gluon initiated fragmentation leads to more suppression of final state particles than quark initiated fragmentation, which is explained by the fact that gluons lose their energy much faster than quarks due to larger color factors in the splitting matrix elements.
Additionally, since the exponent of the momentum fraction $x$ in \eqn{eq:Q-int} is positive ($n-2 >0$) the factor $x^{n-2}$ suppresses the soft modes and the leading parton quenching factor is mostly sensitive to the hard sector constituents which sit at small angle and large momentum fraction. 
\subsection{Jet quenching}
\begin{figure}
    \centering
    \includegraphics[width=0.5\textwidth]{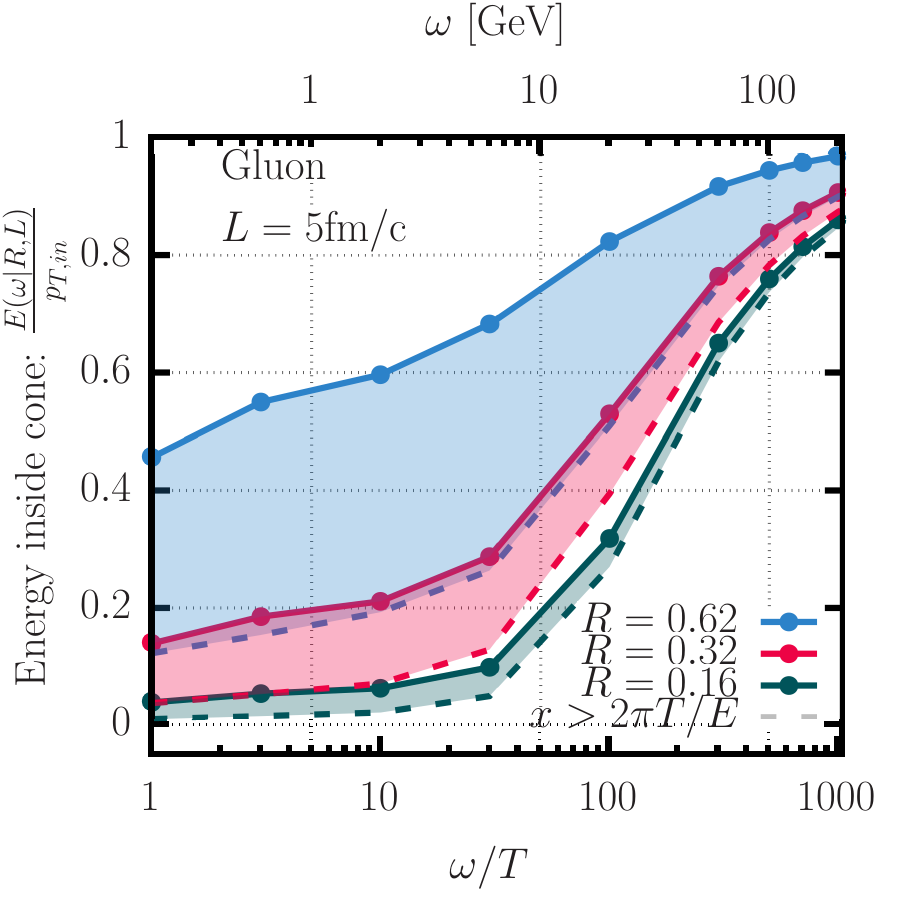}\includegraphics[width=0.5\textwidth]{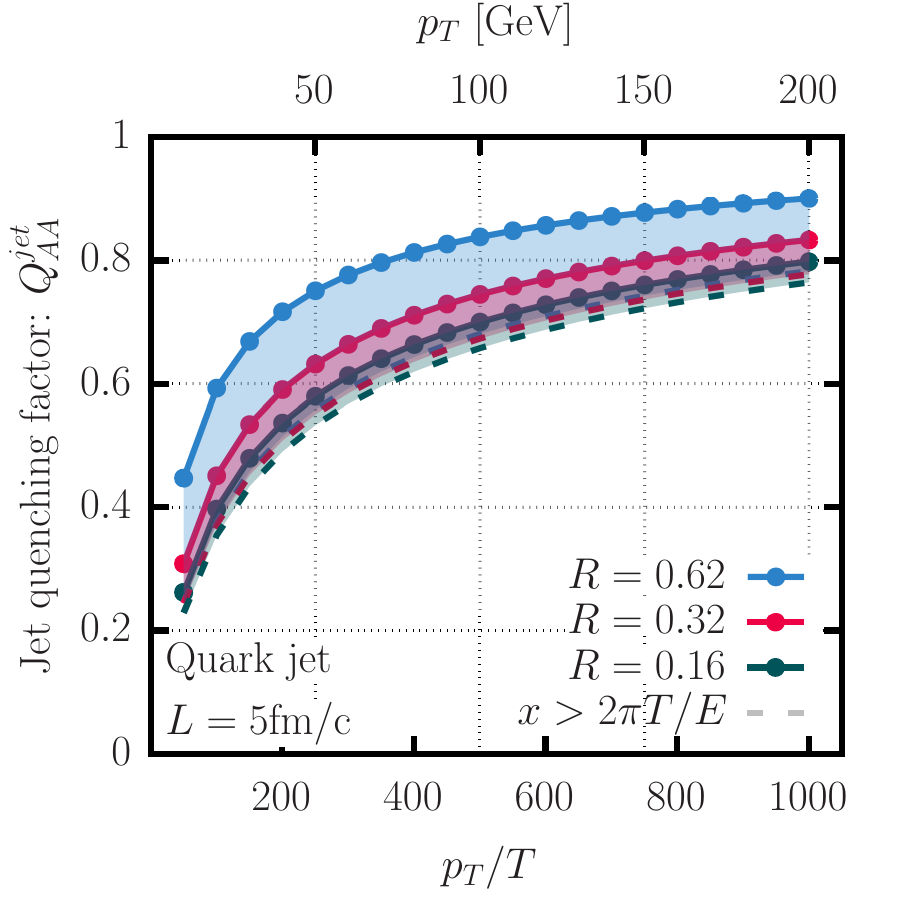}
    \caption{(left) Energy remaining inside the cone for different cone sizes $R=0.16,0.32,0.62$ considering all fragments in full lines or only the hard partons with momentum fraction $x\geq 2\pi T/E$ in dashed lines. 
    (right) Jet quenching factor for a quark jet with cone sizes $R=0.16,0.32,0.62$ after $L=5$fm/c evolution in the medium. Different curves of the same color show the results, including all fragments $x\geq 0$ (solid) or hard fragments only $x\geq 2\pi T/E$ (dashed), such that bands highlight the sensitivity to soft fragments.  
    }
    \label{fig:QuenchingWeight}
\end{figure}

Jets are collimated bunches of particles produced in high energy collisions, which in a first approximation reflect the elementary hard partonic process. The nascent energetic partons created at time scales of order $1/Q$ with large virtuality tend to shower into multi partonic system. This strongly collimated parton cascade is a prediction of pertubative QCD. Of course, beyond these rather generic considerations the precise definition of a jet, from an operational view point, requires the introduction of the notion of the jet cone size $R$ in addition to the use of a jet reconstruction algorithm. 

In the context of heavy ion collisions, it has been recently shown that the collinear in-vacuum parton cascade factorizes from in-medium jet evolution responsible for parton energy loss \cite{Mehtar-Tani:2017web,Mehtar-Tani:2017ypq,Caucal:2018dla,Caucal:2018ofz,Caucal:2019uvr}. The former forms typically over much shorter time scales than medium generated scales. If we consider a single parton splitting, it takes place between $Q^{-1} \equiv (Rp_T)^{-1} \ll t_f  \ll \sqrt{z(1-z) P_T/\hat q }$ where $z$ is the longitudinal momentum fraction carried by the offspring partons. 

In this leading-log picture of jet evolution in the QGP, after the individual color charges that are viewed as single partons by the medium are resolved by it, the in-medium parton cascade discussed in the present manuscript develops. In order to compute the suppression of the inclusive jet spectrum in $AA$ compared to $pp$, one then needs to first compute the probability for a single collinear parton to lose energy out of a jet cone of size $R$.

A full-fledged study of jet quenching withing our kinetic approach is beyond the scope of this work, but we aim in this section to achieve a first step towards providing more computationally friendly tools for jet quenching phenomenology that do not rely on computationally intensive approaches such as MC event generators or transport models. 

Our main goal in this section is therefore to focus on the case of single collinear parton, that is, neglecting in-vacuum jet evolution to gain some insight on how kinetic approaches can be implemented into jet quenching calculations. This will also enable us to address the importance of the medium response of soft sector for inclusive jet spectra.

In the presence of a QGP, the jet spectrum factorizes as follows,

\begin{align}
\frac{\rmd \sigma}{\rmd p_T}= \int_0^\infty \rmd ~\epsilon P(\epsilon,R)\frac{\rmd \sigma_{\rm vac}}{\rmd p_T^{in}}(p_T^{in}\equiv p_T+\epsilon)\;,
\end{align}
where, as before, $p_T^{in}$ denotes the energy of the original (unquenched) parton and $\epsilon$ is the energy loss. Despite the fact that the formula is quite general since the energy loss probability $ P(\epsilon,R)$ may  apply for a multiple partonic system \cite{Mehtar-Tani:2017web} we will restrict our analysis to a single parton.

Following the seminal work by Baier et al. \cite{Baier:2001yt} (see also \cite{Mehtar-Tani:2017web}), because $P(\epsilon,R)$ is dominated by short lived soft gluon radiation that are not measured if they are not considered part of the jet, it takes a Poisson-like form 

\begin{align}
\label{eq:ProbPoission}
 P(\epsilon,R) = \sum_{n=0}^\infty \frac{1}{n!} \left[ \prod_{i=1}^n \omega_i\frac{\rmd I}{\omega_i}\rmd \omega_i \right]  \delta\left(\epsilon -\sum_{i=1}^{n}\omega_i\right) \exp\left[-\int_0^\infty \rmd \omega \frac{\rmd I}{\rmd \omega} \right]
\end{align}

Interestingly, in Laplace space the convolution in $\omega$ space turns into a direct product and hence leads to the exponentiation of the single gluon radiative spectrum 
\begin{align}
 {\cal L}[P(\epsilon,R)](\nu) =\int \rmd \epsilon~{\rm e}^{-\epsilon \nu}  P(\epsilon,R) = \exp\left[- \int \rmd \omega \frac{\rmd I}{\rmd \omega} \left(1-{\rm e}^{  -\omega\nu}\right)\right]
\end{align}
This observation will be used hereafter in evaluating the inclusive jet spectrum. To do so, we need to make an additional approximation. Using the fact that the power jet spectrum is steeply falling, that is, 
\begin{align}
\frac{\rmd \sigma_{\rm vac}}{\rmd p_T}\sim \frac{1}{p_T^n} \quad \text{with} \quad n \gg 1\,,
\end{align}
one may thus expand the shifted power spectrum $(P_T+\epsilon)^{-n}$ around small $\epsilon \ll p_T$ while keeping $ n\epsilon$ finite. This yields, 
\begin{align}
\frac{1}{(p_T+\epsilon)^n } = \frac{1}{p_T^n } \exp\left[ - \frac{n \epsilon}{p_T}\right] \left(1+ {\cal O} (n \epsilon^2)\right)
\end{align}
This is the exponential we need to connect with the Laplace transform. Thus, we have
\begin{align}
\frac{\rmd \sigma}{\rmd p_T}=Q(p_T,R)\frac{\rmd \sigma_{\rm vac}}{\rmd p_T}\,.
\end{align}
where the quenching weight $Q$ is simply the Laplace transform of the energy loss probability evaluated at $\nu=n/p_T$,
\begin{align}\label{eq:QuenchingWeight}
Q(p_T,R) \approx   {\cal L}[P(\epsilon,R)](n/p_T) = \exp\left[- \int \rmd \omega \frac{\rmd I}{\rmd \omega} \left(1-{\rm e}^{ -n\omega/p_T}\right)\right]
\end{align}

Since Eq.~(\ref{eq:ProbPoission}) accounts for multiple emissions in a probabilistic fashion, this formula has the advantage of accounting for the fluctuations of energy loss. However, it neglects the possibility that part of the radiated energy may remain inside the jet cone, hence contributing to the total jet energy. On the other hand, the kinetic equation, while it allows to treat the full dynamics of energy loss down to the dissipation scale, cannot be directly used to compute jet quenching as it allows to compute the inclusive average parton distribution and hence, does not handle fluctuations in and out of the jet cone. However, we may combine the strengths of these two approaches by accounting for fluctuations of energy loss via the Poisson like distribution of the primary gluon emissions and by assuming that after being radiated the primary gluons will rapidly cascade in to a large number of gluons allowing for an event by event averaging of energy in and out of cone.

By replacing 
\begin{align}
\omega \to \omega-E(\omega|R,\tau_{L})
\end{align}
in Eq.~(\ref{eq:QuenchingWeight}), this assumption amounts to shifting the frequency of the primary radiated gluon $\omega$ by the average energy $E(\omega|R,\tau_{L})$ remaining inside a cone of size $R$ after evolving in the medium for a time $\tau_{L}$, which we compute using kinetic theory as in Eq.~(\ref{eq:FullCone}). This idea was previously explored in a simplified fashion in \cite{Mehtar-Tani:2021fud}. 



Now following this approach, let us consider an initial hard quark with energy $p_T$, radiating the first gluon emission with energy $\omega$ along the $z$-axis following the finite-size radiation rate in the soft radiation $\omega$ limit written as \cite{Arnold:2008iy}
\begin{align}
\frac{\rmd\rm I^{q}_{gq}}{\rmd \omega \rmd t} \equiv \frac{ {\rm d} \Gamma^{q}_{gq} }{ {\rm d}  \omega} (p_T,\omega) =
 \frac{\alpha_s}{2\pi} C_{F} \sqrt{\frac{C_A\hat{\bar{q}}(p_T)}{\omega^3}}
\end{align}
and we will employ $\hat{\bar{q}}(p_T)=787\, {\rm MeV/fm}^2$ in the following numerical estimates.
Once it is emitted after a time $t$, the gluon traverses the medium and loses its energy following the kinetic evolution, leaving the medium after time $L$. Its energy is then suppressed by the medium during a time $(L-t)$, therefore the energy remaining in the cone of size $R$ is given by the energy fraction computed in Sec.~\ref{sec:EnergyLoss} evaluated at $(L-t)$, which leads to the slightly modified function $E\left(\omega| R,\tau=\frac{L-t}{t_{\rm th}} \right)$ from Eq.~(\ref{eq:FullCone}) where we include a variable $\omega$ to account for the starting energy of the gluon in the kinetic evolution. We have computed a kinetic evolution for the energies $\omega=(1,3,10,30,100,300,500,700,1000)T$ which we interpolate between to obtain the energy remaining inside the cone as a smooth function of the initial gluon energy $\omega$, making use of the time scaling with the stopping time $t_{\rm th}$ which works fairly well for small cone sizes. 
Using the energy remaining inside the cone after a time $(L-t)$, the quenching weight in Eq.~(\ref{eq:QuenchingWeight}) is then written as
\begin{eqnarray}
Q(p_T) = \exp\left[\int_{0}^{L} {\rm d} t \int {\rm d} \omega \frac{ {\rm d} \Gamma}{ {\rm d}  \omega}\left(1- {\rm e}^{ -n \frac{\omega}{p_T}\left[1-\frac{E\left(\omega| R,\tau=\frac{L-t}{t_{\rm th}} \right)}{\omega}\right] }\right)\right]\;.
\end{eqnarray}
In order to investigate the sensitivity to the soft fragments, we also employ the modified energy inside the cone $E_{\rm hard}\left(\omega| R,\tau=\frac{L-t}{t_{\rm th}} \right)$ from Eq.~(\ref{eq:HighCone}) which account for only the hard fragments with momentum fraction $x\geq 2\pi T/E$. 
In the left panel of Fig.~\ref{fig:QuenchingWeight}, we display the remaining energy inside a cone of size $R$ as a function of the initial parton momentum $p_T$ at $L=5$fm/c for the different cone-sizes $R=0.16,0.32,0.62$. Full line-points represent the full energy $E\left(\omega|R,\tau=\frac{L}{t_{\rm th}} \right)$, while dashed lines represent $E_{\rm hard}\left(\omega| R,\tau=\frac{L}{t_{\rm th}} \right)$, i.e., only the energy stored in the hard partons with momentum fraction $x\geq 2\pi T/E$. The right panel of Fig.~\ref{fig:QuenchingWeight} shows the jet quenching factor for a single prong quark jet, i.e. a quark initiated jet loosing energy as one initial hard quark, using the same color coding. We observe that by passing through a medium of length $L=5 {\rm fm}/c$, single prong quark jets with energies $p_T\leq 25$GeV have lost a significant fraction of their energy out to large cones $R\geq 0.62$.
While the jet quenching factor shows a large suppression of at low $p_T$, depending on the cone size the suppression becomes milder for $p_T \geq 200-400$T. When comparing the results for the different cone-sizes, we find that while for large cone-sizes a clear sensitivity to the soft fragments is apparent, for a narrow cone, the contribution of soft fragments to the energy inside the cone is negligible leading to negligible difference in the quenching. We obtain a clear indication that jet cone sizes larger $\gtrsim 0.3$ are more sensitive to the soft fragments which heavily depend on the details of the thermalization of the soft sector, thus requiring more refined studies of near-equilibrium physics and recoils of the jet onto the medium.

\section{Conclusion}\label{sec:Conclusion}
In this paper, we investigated the dynamics of energy loss and thermalization of a hard parton as it passes through a hot deconfined QCD plasma.
As the highly energetic parton traverses a quark-gluon plasma, it undergoes interactions with the medium, leading to a transfer of energy from the hard modes $(\sim E\ll T)$ all the way to near-equilibrium $(\sim T)$ soft modes.
While the hard modes can be treated perturbatively, this treatment starts to break down once the energy becomes soft and is deposited in the medium, similarly for the response of the medium due to energy loss. Whereas, typically, studies of the thermalization transform the energy deposited into a source term for the hydrodynamic simulation \cite{Casalderrey-Solana:2004fdk,Renk:2006mv,Qin:2009uh,Li:2010ts,Tachibana:2014lja,Tachibana:2017syd,Casalderrey-Solana:2020rsj,Yang:2022nei}, during this study we instead used an effective kinetic description to bridge the apparent gap in the theoretical description of jet quenching. By using a common framework for the hard parton and the soft fragments, we are able to follow the hard parton's energy as it is carried out to large angles and thermalizes with the medium.  

We employed a leading order effective kinetic description of the interactions of a hard parton with a thermal medium, which takes into account both the elastic interactions with the medium and the collinear medium-induced radiation. 
Several studies have shown that the medium cascade is governed by the successive splitting driving an inverse energy cascade \cite{Blaizot:2013hx,Blaizot:2015jea,Mehtar-Tani:2018zba,Schlichting:2020lef}. Since the inelastic radiation is enhanced in the collinear region, this energy cascade is collinear to the leading hard parton. Conversely, the elastic interactions are suppressed at large energies, such that the broadening of the hard parton is marginal. We observe only a small diffusion of the energy, which dies as a fourth power of the angle due to the Moli\`ere tail. 
However, the elastic interactions play a more important role for the soft scales, where the scattering rate becomes comparable to the inelastic interactions. Therefore, once the energy is transferred to the low energy modes, elastic scatterings thermalizes the partons with the medium and push the distribution out to large angles.    



While our study has been centered around the energy loss of a hard parton once it has already been resolved by the medium, we also illustrated how kinetic theory can be used in phenomenological studies of parton/jet quenching. In the future, it would be interesting to further refine the effective kinetic description, by considering more realistic initial conditions that emerge from the early vaccuum-like emissions and lifting some assumptions that simplify the dynamics. By considering the parton to spend a sufficiently long time inside the medium, we employed an infinite medium calculation to obtain a simpler form of the effective medium-induced rates that does not explicitly depend on time. However, for realistic medium sizes, this simplification also leads to a significant over-quenching of partons and jets.
Clearly, a more realistic description can readily be obtained by employing a time dependent splitting rate that captures the interplay of the formation time of the radiation and the medium length  \cite{CaronHuot:2010bp,Andres:2020vxs,Schlichting:2021idr}. When one eventually implements these finite length effects, the simulation can also move away from the simple description of the medium background as a static equilibrated medium and allow for a hydrodynamic evolution of the background QGP medium.

\section{Acknowledgements}
S.S. and I.S. are supported by the Deutsche Forschungsgemeinschaft (DFG, German Research Foundation) through the CRC-TR 211 ’Strong-interaction matter under extreme conditions’– project number 315477589 – TRR 211 and the German Bundesministerium f\"{u}r Bildung und
Forschung (BMBF) through Grant No. 05P21PBCAA. 
I.S. is supported in part by the U.S. Department of Energy (DOE) under grant number DE-SC0013460 and in part by the National Science Foundation (NSF) under grant number OAC-2004571 within the framework of the JETSCAPE collaboration.
The authors also gratefully acknowledge computing time provided by the Paderborn Center for Parallel Computing (PC$^2$). This research used resources of the National Energy Research Scientific Computing Center, a DOE Office of Science User Facility supported by the Office of Science of the U.S. Department of Energy under Contract No. DE-AC02-05CH11231.  Y. M.-T. is supported by the U.S. Department of Energy, Office of Science, Office of Nuclear Physics, under contract No. DE- SC0012704 an acknowledges support from the RHIC Physics Fellow Program of the RIKEN BNL Research Center.

\appendix



\section{Comparison with the small angle approximation}\label{ap:Small-Angle}

\begin{figure}
    \centering
    \includegraphics[width=0.5\textwidth]{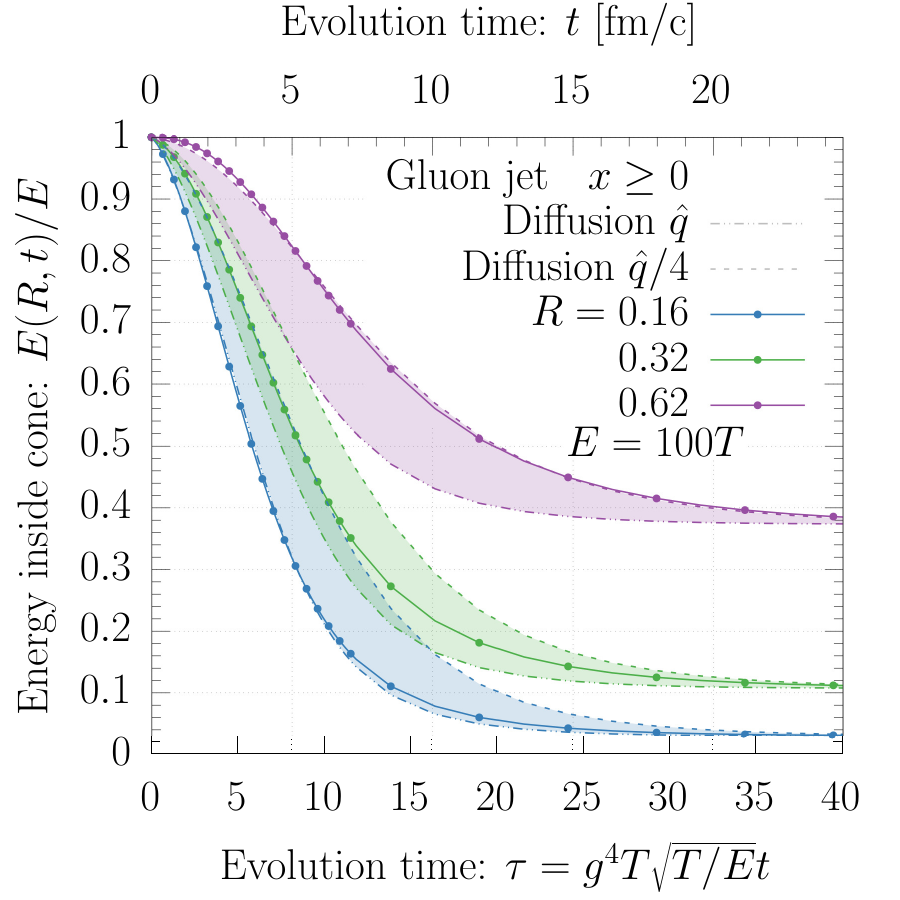}\includegraphics[width=0.5\textwidth]{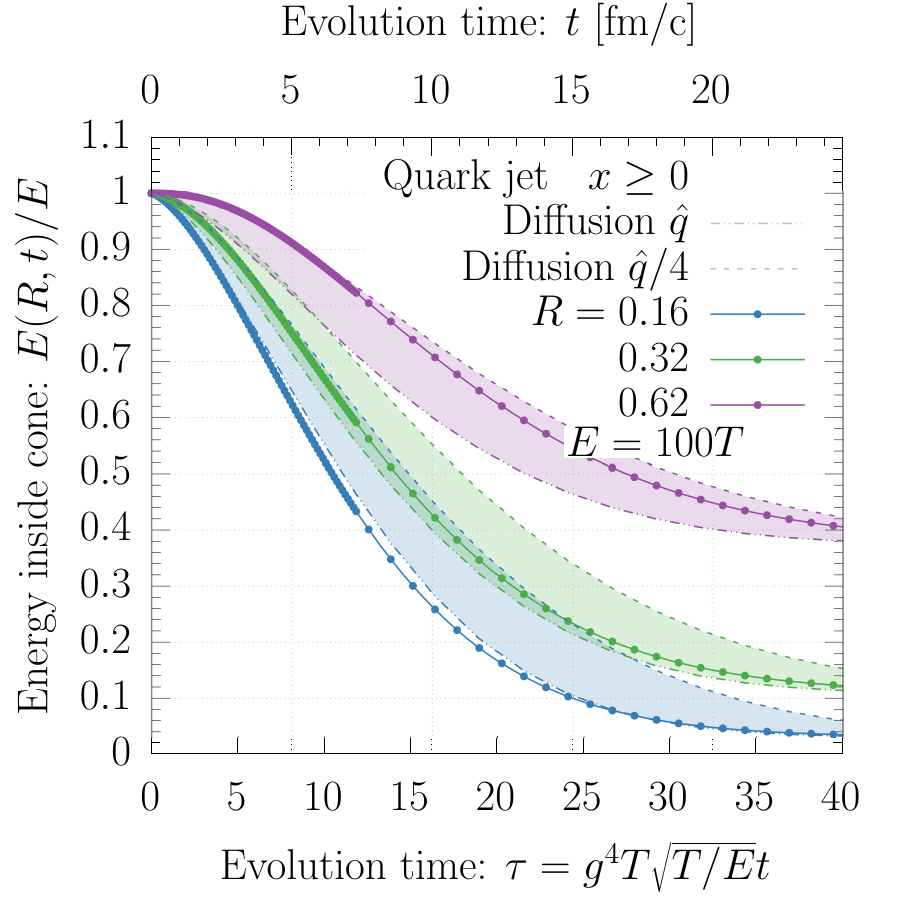}
    \caption{Evolution of the energy inside the cone ($\theta\leq R$) as given by Eq.~(\ref{eq:FullCone}), for gluon (left) and quark (right) initial jets with $E=100T$. We compare with the `Diffusion' approximation, taking the momentum broadening coefficient to be either $\hat{q}$ or $\frac{\hat{q}}{4}$. }
    \label{fig:AngularJetEnergyLossDiff}
\end{figure}

Our investigation of the out-of-cone energy loss in the main text, are based on a leading order QCD kinetic description involving medium induced $1\leftarrow 2$ collinear splittings/mergings and $2\leftarrow2$ elastic interactions described by leading order perturbative matrix elements with HTL screening. While inelastic interactions are responsible for the radiative break-up of hard partons, elastic interactions provide momentum broadening outside of the jet cone. By considering small momentum exchange for elastic interactions the momentum broadening can effectively be described using a Fokker-Planck diffusion equation \cite{Blaizot:2014jna,Ghiglieri:2015ala,Schlichting:2020lef}, where the energy dependence is then accounted for in the Coulomb logarithm of the momentum diffusion coefficient $\hat{q}$. Below we investigate to what extent this `Diffusion' approximation can describe the effects of momentum broadening, both on a qualitative and quantitative level. We chose the equilibrium momentum broadening coefficient $\hat{q}$ for the diffusion evolution where the logarithmic scale dependence is taken to be $\sim 1$,
\begin{equation}\label{eq:Eqqhat}
    \hat{q} = \frac{g^4 T^3 C_R}{2\pi} \left( \frac{N_c}{3}+ \frac{N_f}{6} \right)\;.
\end{equation}
We then proceed as before, and compute the fraction of energy remaining inside the cone ($\theta\leq R$) from Eq.~(\ref{eq:FullCone})
within the full leading order QCD kinetic evolution, and in `Diffusion' approximation. 
The logarithmic dependence of the momentum broadening coefficient, which we do not consider, affects the momentum broadening and consequently out-of-cone energy loss. In the full matrix element evolution this effect is scale dependent, however, in our `Diffusion' simulation we do not set up a scale dependent $\hat{q}$, but instead compare our results with two different values for $\hat{q}$, either taking the equilibrium value in Eq.~(\ref{eq:Eqqhat}) or taking a quarter ($\hat{q}\to\frac{\hat{q}}{4}$). We find that these two values provide a large enough separation to describe the range of cone dependence of the energy loss we focus on in Fig.~\ref{fig:AngularJetEnergyLossDiff}. 
We display in Fig.~\ref{fig:AngularJetEnergyLossDiff} the energy remaining inside the cone $R=(0.16,0.32,0.62)$ from Eq.~(\ref{eq:FullCone}) for the full matrix element evolution using linespoints as well as for the `Diffusion' approximation with long dashed lines for $\hat{q}$ and long dashed lines for $\frac{\hat{q}}{4}$ where on the left we have a Gluon initial jet and on the right a Quark initial jet.
First, we see that, as expected, the evolution using $\frac{\hat{q}}{4}$ loses energy slower than for $\hat{q}$. 
Likewise, how quickly energy is lost also depends on the cone size, where narrower cone sizes lose energy more quickly. 
Comparing with the full matrix element evolution, we observe how the hard particles with narrow cones $R\sim0.16$ are better described by $\hat{q}$ while for larger cone sizes when softer particles are more relevant the evolution is closer to $\frac{\hat{q}}{4}$, confirming our earlier conclusion that, for large cone sizes, the out-of-cone energy loss is sensitive to the detailed dynamics of the soft sector, such that a single scale independent $\hat{q}$ cannot simultaneously describe the broadening at all energy and angular scales.
We note also that the late time limits of the different curves display the energy inside each cone size in the asymptotic distribution, which is the same independently of the evolution used.

\section{Elastic scatterings}\label{ap:HTL}

\begin{table}
	\begin{center}
	\begin{tabular}{c|@{\quad}l@{\quad}}
		$ab \longleftrightarrow cd$ & $\qquad \left|{\cal M}^{ab}_{cd}\right|^2 $ 
		\\[10pt]
		\hline &
		\\[-10pt]
		$
			qq \leftrightarrow qq
		$
		&
		$ \displaystyle
			8 N_f g^4\,  \frac{\df^2 \, \cf^2} {\da} 
			\left( \frac{s^2+u^2}{t^2} + \frac{s^2+t^2}{u^2} \right)
			+
			16 g^4\,  \df \, \cf
			\left( \cf {-} \frac{\ca}{2} \right) \frac{s^2}{tu}
		$
		\\[15pt]
		$q \bar q \longleftrightarrow q \bar q$
		&
		$ \displaystyle
			8N_f g^4\,  \frac{\df^2 \, \cf^2}{\da} 
			\left( \frac{s^2+u^2}{t^2} + \frac{t^2+u^2}{s^2} \right)
			+
			16 g^4\, \df \, \cf
			\left( \cf {-} \frac{\ca}{2} \right) \frac{u^2}{st}
		$
		\\[15pt]
		$q \bar q \longleftrightarrow g \, g$
		&
		$ \displaystyle
			8 g^4\, \df \, \cf^2
			\left( \frac{u}{t} + \frac{t}{u} \right)
			-
			8  g^4\, \df \, \cf \, \ca
			\left( \frac{t^2+u^2}{s^2} \right)
		$
		\\[15pt]
		$
		\begin {array}{c}
			q \, g \longleftrightarrow q \, g \,,\\ \bar q \, g \longleftrightarrow \bar q \, g
		\end {array}
		$
		&
		$ \displaystyle
			-8g^4\, \df \, \cf^2
			\left( \frac{u}{s}  +  \frac{s}{u} \right)
			+
			8g^4\, \df \, \cf \, \ca
			\left( \frac{s^2 + u^2}{t^2} \right)
		$
		\\[15pt]
		$g \, g \leftrightarrow g \, g$
		&
		$ \displaystyle
			16g^4\, \da \, \ca^2
			\left(
			3 - \frac{su}{t^2} - \frac{st}{u^2} - \frac{tu}{s^2}
			\right)
		$
	\end{tabular}
	\end{center}
	\caption{Different $2\leftrightarrow 2 $ processes in QCD and their squared matrix elements \cite{Arnold:2002zm} written in terms of the Mandelstam variables $s=(P_1+P_2)^2$, $t=(P_1-P_3)^2$ and $u=(P_1-P_4)^2$.  For QCD theory with $SU(N_c)$ gauge symmetry, the fundamental representation has dimension $d_F=N_c$ and quadratic Casimir $C_F=\tfrac{N_c^2-1}{2N_c}$, while in the adjoint representation the dimension is $d_A=N_c^2 -1$ and the quadratic Casimir is $C_A=N_c$.   }
	\label{tab:MatElements} 
\end{table}
During this section, we will briefly compile the elastic collision integrals we use during this study.
Since following the evolution for each quark flavor separately is numerically costly, we will instead sum over the different flavor by introducing the distribution, 
\begin{align}
    D_{Q}=\sum_i D_{q_i}\;,\qquad
    D_{\bar{Q}}=\sum_i D_{\bar{q}_i}\;.
\end{align}
We present the elastic scattering matrix element squared in Tab.~\ref{tab:MatElements} in this convention, where processes that can evolve different flavors (e.g., the t-channel of quark-quark scattering) acquire a factor $N_f$. 
The collision kernel of the elastic scattering is written
\begin{align}
    C_a^{2\leftrightarrow 2}[\{f_i\}] = & \frac 1{2|p_1|\nu_a} \sum_{bcd} \int d\Omega^{2\leftrightarrow2}
        \left|{\cal M}^{ab}_{cd}(\p_1,\p_2;\p_3,\p_4)\strut\right|^2 \delta \mathcal{F}(\p_1,\p_2;\p_3,\p_4) \;,
\end{align}
where the statistical factor $\delta\mathcal{F}(\p_1,\p_2,\p_3,\p_4)$ is now 
\begin{align}
    \delta\mathcal{F}(\p_1,\p_2,\p_3,\p_4) =& ~~~\delta f_a(\p_1) \left[ \pm_{a} n_c(p_3)  n_d(p_4) - n_b(p_2)(1 \pm n_c(p_3) \pm n_d(p_4)) \right] \nonumber\\
    &+ \delta f_b(\p_2) \left[ \pm_{b} n_c(p_3)  n_d(p_4) - n_b(p_1)(1 \pm n_c(p_3) \pm n_d(p_4)) \right] \nonumber\\
    &- \delta f_c(\p_3) \left[ \pm_{c} n_a(p_1)  n_b(p_2) - n_b(p_4)(1 \pm n_a(p_1) \pm n_b(p_2)) \right] \nonumber\\
    &- \delta f_d(\p_4) \left[ \pm_{d} n_a(p_1)  n_b(p_2) - n_b(p_3)(1 \pm n_a(p_1) \pm n_b(p_2)) \right] \;,
\end{align}
while $\pm_i$ is plus if particle $i$ is a boson and minus for a fermion.
Since we consider the phase-space distribution to be isotropic in the azimuthal angle $\phi_{p}$ of momentum ${\bf p}$, we can express the phase-space distributions of gluons and quarks as
\begin{equation}
    \delta f_g(x_1,\theta_1) = \int \frac{d\phi_{p}}{2\pi} ~ \delta f_a(\p) = (2\pi)^2 \frac{D_g(x_1,\theta_1)}{\nu_g (x_1 E)^3}\,,\qquad
    \delta f_q(x_1,\theta_1) = (2\pi)^2  \frac{D_Q(x_1,\theta_1)}{N_f\nu_q (x_1 E)^3}
\end{equation}
and similarly for anti-quarks.
Expressing the contributions to the collision kernels in terms of the energy distribution $D_{a}$, the contributions to the gluon collision integral from different processes are given by 
\begin{align}
    C_{g}^{qg\leftrightarrow qg} =&  
        \frac 1{2|p_1|\nu_a}  \int d\Omega^{2\leftrightarrow2}
         \left|{\cal M}^{gq}_{gq}\strut\right|^2 \nonumber\\
        &\Bigg[
         N_f \frac{D_g(x_1,\theta_1)}{\nu_g x_1^3 E^2} \left[ n_F(p_3)  n_B(p_4) - n_b(p_2)(1 - n_F(p_3) + n_B(p_4)) \right] \nonumber\\
        &+ \frac{D_Q(x_2,\theta_2)}{ \nu_q x_2^3 E^2} \left[ - n_F(p_3)  n_B(p_4) - n_F(p_1)(1 - n_F(p_3) + n_B(p_4)) \right] \nonumber\\
        &- \frac{D_Q(x_3,\theta_3)}{\nu_q x_3^3 E^2} \left[ - n_B(p_1)  n_F(p_2) - n_F(p_4)(1 + n_B(p_1) - n_F(p_2)) \right] \nonumber\\
        &- N_f \frac{D_g(x_4,\theta_4)}{\nu_g x_4^3 E^2} 
        \left[  n_B(p_1)  n_F(p_2) - n_F(p_3)(1 + n_B(p_1) - n_F(p_2)) \right] \Bigg] \nonumber\\
        &+\frac 1{2|p_1|\nu_a}  \int d\Omega^{2\leftrightarrow2}
        \left|{\cal M}^{gq}_{qg}\strut\right|^2 \nonumber\\
        &\Bigg[
         N_f \frac{D_g(x_1,\theta_1)}{\nu_g x_1^3 E^2} \left[ n_B(p_3)  n_F(p_4) - n_b(p_2)(1 + n_B(p_3) - n_F(p_4)) \right] \nonumber\\
        &+  \frac{D_Q(x_2,\theta_2)}{\nu_q x_2^3 E^2} \left[ - n_B(p_3)  n_F(p_4) - n_b(p_1)(1 + n_B(p_3) - n_F(p_4)) \right] \nonumber\\
        &- N_f \frac{D_g(x_3,\theta_3)}{ \nu_g x_3^3 E^2} \left[ n_B(p_1)  n_F(p_2) - n_F(p_4)(1 + n_B(p_1) - n_F(p_2)) \right] \nonumber\\
        &- \frac{D_Q(x_4,\theta_4)}{\nu_q x_4^3 E^2} \left[- n_B(p_1)  n_F(p_2) - n_B(p_3)(1 + n_B(p_1) - n_F(p_2)) \right] \Bigg]\;,\\
       C_{g}^{gg\leftrightarrow q\bar{q}} =&\frac 1{2|p_1|\nu_a}  \int d\Omega^{2\leftrightarrow2}
         \left|{\cal M}^{q\bar{q}}_{gg}\strut\right|^2 \nonumber\\
        &\Bigg[
         N_f\frac{D_g(x_1,\theta_1)}{\nu_q x_1^3 E^2} \left[ -n_F(p_3)  n_B(p_4) - n_F(p_2)(1 - n_F(p_3) + n_B(p_4)) \right] \nonumber\\
        &+N_f  \frac{D_g(x_2,\theta_2)}{\nu_q x_2^3 E^2} \left[ n_F(p_3)  n_B(p_4) - n_B(p_1)(1 - n_F(p_3) + n_B(p_4)) \right] \nonumber\\
        &- \frac{D_Q(x_3,\theta_3)}{\nu_q x_3^3 E^2} \left[ - n_F(p_1)  n_B(p_2) - n_F(p_4)(1 - n_F(p_1) + n_B(p_2)) \right] \nonumber\\
        &- \frac{D_{\bar{Q}}(x_4,\theta_4)}{\nu_q x_4^3 E^2} 
        \left[  n_F(p_1)  n_B(p_2) - n_F(p_3)(1 - n_F(p_1) - n_B(p_2)) \right] \Bigg]\;,\nonumber\\
        &+\frac 1{2|p_1|\nu_a}  \int d\Omega^{2\leftrightarrow2}
         \left|{\cal M}^{q\bar{q}}_{gg}\strut\right|^2 \nonumber\\
        &\Bigg[
         N_f\frac{D_g(x_1,\theta_1)}{\nu_q x_1^3 E^2} \left[ -n_F(p_3)  n_B(p_4) - n_F(p_2)(1 - n_F(p_3) + n_B(p_4)) \right] \nonumber\\
        &+N_f  \frac{D_g(x_2,\theta_2)}{\nu_q x_2^3 E^2} \left[ n_F(p_3)  n_B(p_4) - n_B(p_1)(1 - n_F(p_3) + n_B(p_4)) \right] \nonumber\\
        &- \frac{D_{\bar{Q}}(x_3,\theta_3)}{\nu_q x_3^3 E^2} \left[ - n_F(p_1)  n_B(p_2) - n_F(p_4)(1 - n_F(p_1) + n_B(p_2)) \right] \nonumber\\
        &- \frac{D_Q(x_4,\theta_4)}{\nu_q x_4^3 E^2} 
        \left[  n_F(p_1)  n_B(p_2) - n_F(p_3)(1 - n_F(p_1) - n_B(p_2)) \right] \Bigg]\;,\nonumber\\
\end{align}
where the sum over the quark flavors cancels with the factor $1/N_f$ for the quark energy distribution.
Similarly, the contribution of quark-gluon scattering to the quark collision integral, is obtained by performing a sum over the $N_f$ degenerate flavors as follows
\begin{align}
    C_Q^{qg\leftrightarrow qg} =& \sum_f C_{q_f} =  
    \frac 1{2|p_1|\nu_a}  \int d\Omega^{2\leftrightarrow2}
     \left|{\cal M}^{gq}_{gq}\strut\right|^2 \nonumber\\
    &\Bigg[
     \frac{D_Q(x_1,\theta_1)}{\nu_q x_1^3 E^2} \left[ -n_F(p_3)  n_B(p_4) - n_F(p_2)(1 - n_F(p_3) + n_B(p_4)) \right] \nonumber\\
    &+  N_f \frac{D_g(x_2,\theta_2)}{\nu_g x_2^3 E^2} \left[ n_F(p_3)  n_B(p_4) - n_B(p_1)(1 - n_F(p_3) + n_B(p_4)) \right] \nonumber\\
    &- \frac{D_Q(x_3,\theta_3)}{ \nu_q x_3^3 E^2} \left[ - n_F(p_1)  n_B(p_2) - n_F(p_4)(1 - n_F(p_1) + n_B(p_2)) \right] \nonumber\\
    &- N_f \frac{D_g(x_4,\theta_4)}{\nu_g x_4^3 E^2} 
    \left[  n_F(p_1)  n_B(p_2) - n_F(p_3)(1 - n_F(p_1) - n_B(p_2)) \right] \Bigg] \nonumber\\
    &+\frac 1{2|p_1|\nu_a}  \int d\Omega^{2\leftrightarrow2}
    N_f \left|{\cal M}^{gq}_{qg}\strut\right|^2 \nonumber\\
    &\Bigg[
     \frac{D_Q(x_1,\theta_1)}{\nu_q x_1^3 E^2} \left[ n_B(p_3)  n_F(p_4) - n_B(p_2)(1 + n_B(p_3) - n_F(p_4)) \right] \nonumber\\
    &+  N_f \frac{D_g(x_2,\theta_2)}{\nu_g x_2^3 E^2} \left[ - n_B(p_3)  n_F(p_4) - n_F(p_1)(1 + n_B(p_3) - n_F(p_4)) \right] \nonumber\\
    &- N_f \frac{D_g(x_3,\theta_3)}{\nu_g x_3^3 E^2} \left[ n_F(p_1)  n_B(p_2) - n_F(p_4)(1 - n_F(p_1) + n_B(p_2)) \right] \nonumber\\
    &- \frac{D_Q(x_4,\theta_4)}{\nu_q x_4^3 E^2} \left[- n_B(p_1)  n_b(p_2) - n_B(p_3)(1 - n_F(p_1) + n_B(p_2)) \right] \Bigg]\;.
\end{align}
Next, the contributions of elastic scattering of quarks is given by 
\begin{align}
    C_Q^{qq\leftrightarrow qq} =& \sum_i C_{q_i} =  
    \frac 1{2|p_1|\nu_a}  \int d\Omega^{2\leftrightarrow2}
     \left|{\cal M}^{qq}_{qq}\strut\right|^2 \nonumber\\
    &\Bigg[
     \frac{D_Q(x_1,\theta_1)}{\nu_q x_1^3 E^2} \left[ -n_F(p_3)  n_B(p_4) - n_F(p_2)(1 - n_F(p_3) + n_B(p_4)) \right] \nonumber\\
    &+  \frac{D_Q(x_2,\theta_2)}{\nu_q x_2^3 E^2} \left[ n_F(p_3)  n_B(p_4) - n_B(p_1)(1 - n_F(p_3) + n_B(p_4)) \right] \nonumber\\
    &- \frac{D_Q(x_3,\theta_3)}{\nu_q x_3^3 E^2} \left[ - n_F(p_1)  n_B(p_2) - n_F(p_4)(1 - n_F(p_1) + n_B(p_2)) \right] \nonumber\\
    &- \frac{D_Q(x_4,\theta_4)}{\nu_q x_4^3 E^2} 
    \left[  n_F(p_1)  n_B(p_2) - n_F(p_3)(1 - n_F(p_1) - n_B(p_2)) \right] \Bigg]\;,
\end{align}
where we combined the two processes $q_iq_f \to q_i q_f$ and $q_i q_f \to q_f q_i$ by writing the second processes as u-channel matrix element, and added the process $q_i q_i \to q_i q_i$ in order to obtain the full matrix element in Tab.\ref{tab:MatElements}.

Eventually, the remaining quark anti-quark scattering and annihilation processes are written  as
\begin{align}
    C_Q^{q\bar{q}\leftrightarrow q\bar{q}} =& \sum_i C_{q_i} =  
    \frac 1{2|p_1|\nu_a}  \int d\Omega^{2\leftrightarrow2}
     \left|{\cal M}^{q\bar{q}}_{q\bar{q}}\strut\right|^2 \nonumber\\
    &\Bigg[
     \frac{D_Q(x_1,\theta_1)}{\nu_q x_1^3 E^2} \left[ -n_F(p_3)  n_B(p_4) - n_F(p_2)(1 - n_F(p_3) + n_B(p_4)) \right] \nonumber\\
    &+  \frac{D_{\bar{Q}}(x_2,\theta_2)}{\nu_q x_2^3 E^2} \left[ n_F(p_3)  n_B(p_4) - n_B(p_1)(1 - n_F(p_3) + n_B(p_4)) \right] \nonumber\\
    &- \frac{D_Q(x_3,\theta_3)}{\nu_q x_3^3 E^2} \left[ - n_F(p_1)  n_B(p_2) - n_F(p_4)(1 - n_F(p_1) + n_B(p_2)) \right] \nonumber\\
    &- \frac{D_{\bar{Q}}(x_4,\theta_4)}{\nu_q x_4^3 E^2} 
    \left[  n_F(p_1)  n_B(p_2) - n_F(p_3)(1 - n_F(p_1) - n_B(p_2)) \right] \Bigg]\;,\nonumber\\
    &+\frac 1{2|p_1|\nu_a}  \int d\Omega^{2\leftrightarrow2}
     \left|{\cal M}^{q\bar{q}}_{\bar{q}q}\strut\right|^2 \nonumber\\
    &\Bigg[
     \frac{D_Q(x_1,\theta_1)}{\nu_q x_1^3 E^2} \left[ -n_F(p_3)  n_B(p_4) - n_F(p_2)(1 - n_F(p_3) + n_B(p_4)) \right] \nonumber\\
    &+  \frac{D_{\bar{Q}}(x_2,\theta_2)}{\nu_q x_2^3 E^2} \left[ n_F(p_3)  n_B(p_4) - n_B(p_1)(1 - n_F(p_3) + n_B(p_4)) \right] \nonumber\\
    &- \frac{D_{\bar{Q}}(x_3,\theta_3)}{\nu_q x_3^3 E^2} \left[ - n_F(p_1)  n_B(p_2) - n_F(p_4)(1 - n_F(p_1) + n_B(p_2)) \right] \nonumber\\
    &- \frac{D_Q(x_4,\theta_4)}{\nu_q x_4^3 E^2} 
    \left[  n_F(p_1)  n_B(p_2) - n_F(p_3)(1 - n_F(p_1) - n_B(p_2)) \right] \Bigg]\;,\\
    C_Q^{q\bar{q}\leftrightarrow gg} =& \sum_i C_{q_i} =  
    \frac 1{2|p_1|\nu_a}  \int d\Omega^{2\leftrightarrow2}
     \left|{\cal M}^{q\bar{q}}_{gg}\strut\right|^2 \nonumber\\
    &\Bigg[
     \frac{D_Q(x_1,\theta_1)}{\nu_q x_1^3 E^2} \left[ -n_F(p_3)  n_B(p_4) - n_F(p_2)(1 - n_F(p_3) + n_B(p_4)) \right] \nonumber\\
    &+  \frac{D_{\bar{Q}}(x_2,\theta_2)}{\nu_q x_2^3 E^2} \left[ n_F(p_3)  n_B(p_4) - n_B(p_1)(1 - n_F(p_3) + n_B(p_4)) \right] \nonumber\\
    &- N_f\frac{D_g(x_3,\theta_3)}{\nu_q x_3^3 E^2} \left[ - n_F(p_1)  n_B(p_2) - n_F(p_4)(1 - n_F(p_1) + n_B(p_2)) \right] \nonumber\\
    &- N_f\frac{D_g(x_4,\theta_4)}{\nu_q x_4^3 E^2} 
    \left[  n_F(p_1)  n_B(p_2) - n_F(p_3)(1 - n_F(p_1) - n_B(p_2)) \right] \Bigg]\;,\nonumber\\
\end{align}
where, we combined the two processes $q_i\bar{q}_i \to q_i \bar{q}_i$ and $q_i \bar{q}_i \to \bar{q}_f q_f$.
Analogously, one can obtain the remaining processes involving anti-quark particles.
In the following sections, we will outline our treatment of the matrix element using Hard thermal loop (HTL) propagators.
\subsection{Hard thermal loop matrix element}

The matrix elements in Tab.~\ref{tab:MatElements} are for particles in vacuum, which leads to an infrared divergence in the momentum exchange in the t- and u-channels\footnote{The mixed channels $s^2/(tu)$ and $u^2/(st)$ also generate divergences, but these cancel between the gain and loss term \cite{Arnold:2002zm}.}. For a proper treatment using thermal propagators, the divergences are regulated by the self-energies. However, as the authors of \cite{Arnold:2002zm} argue, medium-dependent effects are only important for small angle scatterings corresponding to the regions where $-t$ or $-u$ are of the order of the thermal mass squared and the modification of the other terms can be neglected.
In this section, we will follow the AMY approach \cite{Arnold:2002zm,Arnold:2002ja,Arnold:2003zc} in order to re-write the divergent terms using the retarded self-energy, which cuts off the divergence. 

\paragraph{Gluon exchange:}
We apply the rewriting for the following t- and u-channel gluon exchange processes: $gg \leftrightarrow gg$, $q^i q^j \leftrightarrow q^i q^j $, $q^i \bar{q}^j \leftrightarrow q^i \bar{q}^j $, $\bar{q}^i \bar{q}^j \leftrightarrow \bar{q}^i \bar{q}^j $ and $q^i  g \leftrightarrow q^i g$.
The matrix elements in terms of Mandelstam variables can be rewritten as
\begin{align}
    \frac{s^2+u^2}{t^2} = \frac{1}{2} + \frac{1}{2} \frac{(s-u)^2}{t^2}\;,&\qquad
    \frac{su}{t^2} = \frac{1}{4} - \frac{1}{4} \frac{(s-u)^2}{t^2}\;,\\
    \frac{s^2+t^2}{u^2} = \frac{1}{2} + \frac{1}{2} \frac{(s-t)^2}{u^2}\;,&\qquad
    \frac{st}{u^2} = \frac{1}{4} - \frac{1}{4} \frac{(s-t)^2}{u^2}\;.
\end{align}
Using thermal propagators amounts to the replacement \cite{Arnold:2003zc}
\begin{align}
    \frac{(s-u)^2}{t^2} \longrightarrow |G_{\mu\nu}(P_1- P_3) (P_1+P_3)^\mu (P_2+P_4)^\nu|^2\;, \\
    \frac{(s-t)^2}{u^2} \longrightarrow |G_{\mu\nu}(P_1- P_4) (P_1+P_4)^\mu (P_2+P_3)^\nu|^2\;,
\end{align}
where $G_{\mu\nu}(P_1- P_3)$ is the retarded thermal gluon propagator, computed in the HTL approximation. 
In the Coulomb gauge, it is given by
\begin{align}
    G_{00}(\omega,\q) =& \frac{-1}{q^2 + \Pi_{00}(\omega,q)}\;,\qquad
    G_{ij}(\omega,\q) = \frac{\delta_{ij} - \tfrac{\q_i\q_j}{q^2}}{q^2 - \omega^2 + \Pi_{T}(\omega,q)}\;,\\
    G_{i0}(\omega,\q) =& G_{0i}(\omega,\q) = 0\;.
\end{align}
The transverse and longitudinal gluon self-energies are given by 
\begin{align}
    \Pi_{00}(\omega,q) =& m_D^2 \left[ 1 - \frac{\omega }{2q} \left( \ln \left( \frac{q+\omega}{q-\omega}\right) - i\pi \right) \right]\;,\\
    \Pi_{T}(\omega,q) =& m_D^2 \left[ \frac{\omega^2}{q^2} + \frac{\omega (q^2-\omega^2)}{4q^3} \left( \ln \left( \frac{q+\omega}{q-\omega}\right) - i\pi \right) \right]\;,
\end{align}
in the $(+---)$ metric convention, the propagator with 4-momentum $Q=(\omega,\q)$ is space-like (i.e., $|\omega|<q$) for the t- and u-channels and the logarithm is well-behaved. 
\paragraph{Quark exchange:}
The other t- and u-channel processes that require a rewriting are the quark exchange processes: $q^i \bar{q}^i \leftrightarrow gg $ and $q g \leftrightarrow qg$.
The matrix elements are computed using the following replacement for the four-momentum exchange,
\begin{equation}
    Q^{\mu} \to \mathcal{Q}^\mu \equiv Q^\mu - \Sigma^{\mu}(Q)\;,
\end{equation}
where the quark self-energy in the Coulomb gauge is given by
\begin{align}
    \Sigma^0(Q) =& \frac{m_F^2}{2q} \left[ \ln\left( \frac{q+\omega}{q-\omega}\right) -i\pi \right]\;,\\
    \bm\Sigma(Q) =& -\q \frac{m_F^2}{q^2} \left[ 1-\frac{\omega}{2q} \left( \ln\left( \frac{q+\omega}{q-\omega}\right) -i\pi \right) \right]\;.
\end{align}
The relevant matrix elements become 
\begin{align}
    \frac{u}{t} \longrightarrow& \left. \frac{4 {\rm Re} [ (P_1\cdot\mathcal{Q})(P_2\cdot\mathcal{Q}^*)] - s \mathcal{Q}\cdot\mathcal{Q}^*}{|\mathcal{Q}\cdot\mathcal{Q}|^2} \right|_{\mathcal{Q}^\mu = P_1^\mu -P_3^\mu  - \Sigma^{\mu}(P_1 - P_3)}\;,\\
    \frac{t}{u} \longrightarrow& \left. \frac{4 {\rm Re} [ (P_1\cdot\mathcal{Q})(P_2\cdot\mathcal{Q}^*)] - s \mathcal{Q}\cdot\mathcal{Q}^*}{|\mathcal{Q}\cdot\mathcal{Q}|^2} \right|_{\mathcal{Q}^\mu = P_1^\mu -P_4^\mu  - \Sigma^{\mu}(P_1 - P_4)}\;,\\
    \frac{s}{u} \longrightarrow& \left. \frac{-4 {\rm Re} [ (P_1\cdot\mathcal{Q})(P_3\cdot\mathcal{Q}^*)] - t \mathcal{Q}\cdot\mathcal{Q}^*}{|\mathcal{Q}\cdot\mathcal{Q}|^2} \right|_{\mathcal{Q}^\mu = P_1^\mu -P_4^\mu  - \Sigma^{\mu}(P_1 - P_4)}\;.
\end{align}





\section{Asymptotic distributions}\label{ap:Equilibirum}
Below we derive the asymptotic distributions for the energy distribution $D_{a}(x,\theta)$. We first consider the change of the volume integrated phase-space distribution
\beq 
\delta \bar{f}^{(eq)}_{a}({\bf p})= \int _{{\bf x}} \delta \bar{f}^{(eq)}_{a}({\bf x},{\bf p}) 
\eeq
which can be represented as a linear superposition of the changes of the temperature $T$, the rest-frame four velocity $u^{z}$ and the chemical potential $\frac{\mu_{f}}{T}$ according to
\beq
\label{eq:fbarpert}
\delta \bar{f}^{(eq)}_{a}({\bf p})= V \left[ \frac{\delta T}{T} T\partial_{T} + \delta u^{z} \partial _{u^{z}} + \delta \Big(\frac{\mu_{f}}{T} \Big) \partial_{\frac{\mu_{f}}{T}} \right] n_{a}({\bf p})\;.
\eeq
We note that in the above expression the factor of the volume $V$ has to appear on dimensional grounds, however this will cancel against the amplitudes $\delta T/T$, $\delta u^{z}$ and $\delta \Big(\frac{\mu_{f}}{T} \Big)$ all of which are inversely proportional to the volume. Specifically, by performing the thermodynamic matching for an ideal gas of massless partons, these amplitudes can be expressed as
\beq 
\label{eq:deltaAmplitudes}
\frac{\delta T}{T} &=& \frac{1}{4} \frac{\delta E}{E_{\rm tot}} = \frac{1}{4}\frac{\delta E}{V e(T)}\;, \\
\delta u^{z}&=&\frac{\delta P^{z}}{E_{\rm tot}+P_{\rm tot}}=\frac{3}{4} \frac{\delta P^{z}}{V e(T)}\;, \\
\delta \Big(\frac{\mu_f}{T} \Big) &=& \frac{\delta N_f}{V \chi(T) T},
\eeq 
where $E_{\rm tot}=V e(T)$ and $P_{\rm tot}=E_{\rm tot}/3$ are the total energy and total pressure, $e(T)=\left( \frac{\pi^2}{30} \nu_g + \frac{7 \pi^2}{120} \nu_q N_{f} \right)T^4$ is the energy density, and $\chi(T)=\frac{\nu_q}{6} T^2$ is the charge susceptibility. By inserting Eqns.~(\ref{eq:fbarpert}) and (\ref{eq:deltaAmplitudes}) into the definition of the energy distribution in Eq.~(\ref{eq:FragmentationFct}) and employing $\delta E=\delta P_z$ equal to the energy $E$ of the jet, the asymptotic energy distribution then takes the form
\beq
D_{a}^{(eq)}(x,\theta)=\frac{\nu_a}{(2\pi)^2}(xE)^3\left[ \frac{1}{4} \frac{E}{e(T)} T \partial_T + \frac{3}{4} \frac{P^{z}}{e(T)} \partial _{u^{z}} + \frac{\delta N_f}{\chi(T)T}  \partial_{\frac{\mu_{f}}{T}} \right] \left.n_{a}\Big(p_{\mu}u^{\mu} \Big) \right|_{u^{z}=0}
\eeq
which is to be
evaluated at $|\mathbf{p}|=xE$, with $\delta N_{f}=0$ for gluon jets and $\delta N_{f}=\pm \delta_{fi}$ for quark/anti-quark jets of a given flavor $i$. Since the change due to the chemical potential contributes with opposite signs for particles and anti-particles, it cancels in the total energy distribution, which is then given by
\beq
\label{eq:DeqSol}
D^{(eq)}(x,\theta)= \sum_{a}
\frac{\nu_a}{(2\pi)^2} \frac{(xE)^4}{4 e(T)}~\frac{E}{T}~\left[ 1+3\cos(\theta)\right]n_{a}(xE) (1\pm n_{a}(xE))
\eeq
where in the last step we used the identity
\beq
 T \partial_{T} n_{a}(p) = -p \partial_{p} n_{a}(p)=  \frac{p}{T} n_{a}(p) \Big( 1\pm n_{a}(p)\Big)\;.
\eeq
By use of this identity it is also straightforward to verify that the asymptotic distribution in Eq.~(\ref{eq:DeqSol}) satisfies the normalization condition in Eq.~(\ref{eq:NormD}).
\bibliography{main.bib} 
\bibliographystyle{ieeetr}

\end{document}